\documentclass[a4paper, amsfonts, amssymb, amsmath, reprint, showkeys, nofootinbib, superscriptaddress]{revtex4-1}
\usepackage[english]{babel}
\usepackage[utf8]{inputenc}
\usepackage[colorinlistoftodos, color=green!40, prependcaption]{todonotes}
\usepackage{amsthm}
\usepackage{mathtools}
\usepackage{physics}
\usepackage{mhchem}
\usepackage{xcolor}
\usepackage{graphicx}
\usepackage[margin=18mm,top=35mm,columnsep=15pt]{geometry} 
\usepackage{adjustbox}
\usepackage{float}
\usepackage{placeins}
\usepackage[T1]{fontenc}
\usepackage{lipsum}
\usepackage{csquotes}

\makeatletter
\def\fnum@figure{FIG.~\thefigure}
\makeatother
\usepackage[pdftex, pdftitle={Article}, pdfauthor={Author}]{hyperref} % For hyperlinks in the PDF
\newcommand{\nocontentsline}[3]{}
\let\origcontentsline\addcontentsline
\newcommand\stoptoc{\let\addcontentsline\nocontentsline}
\newcommand\resumetoc{\let\addcontentsline\origcontentsline}

\begin{document}

\title{All-Optical Photonic Crystal Bolometer with Ultra-Low Heat Capacity for Scalable Thermal Imaging}

\author{Louis Follet}
\email[]{louisf13@mit.edu}
\affiliation{Research Laboratory of Electronics, Massachusetts Institute of Technology, Cambridge, MA 02139, USA}
% \affiliation{Electrical Engineering and Computer Science, Massachusetts Institute of Technology, Cambridge, MA 02139, USA}

\author{Jordan Goldstein}
\affiliation{Research Laboratory of Electronics, Massachusetts Institute of Technology, Cambridge, MA 02139, USA}
% \affiliation{Electrical Engineering and Computer Science, Massachusetts Institute of Technology, Cambridge, MA 02139, USA}

\author{Christopher L. Panuski}
\affiliation{Research Laboratory of Electronics, Massachusetts Institute of Technology, Cambridge, MA 02139, USA}
% \affiliation{Electrical Engineering and Computer Science, Massachusetts Institute of Technology, Cambridge, MA 02139, USA}

\author{Ian Christen}
\affiliation{Research Laboratory of Electronics, Massachusetts Institute of Technology, Cambridge, MA 02139, USA}
% \affiliation{Electrical Engineering and Computer Science, Massachusetts Institute of Technology, Cambridge, MA 02139, USA}

\author{Sivan Trajtenberg-Mills}
\affiliation{School of Electrical Engineering, Faculty of Engineering, Tel Aviv University, Tel Aviv 6997801, Israel}

\author{Dirk R. Englund}
\email[]{englund@mit.edu}
\affiliation{Research Laboratory of Electronics, Massachusetts Institute of Technology, Cambridge, MA 02139, USA}
% \affiliation{Electrical Engineering and Computer Science, Massachusetts Institute of Technology, Cambridge, MA 02139, USA}

% \date{\today} % Leave empty to omit a date

\begin{abstract}
High-speed thermal imaging in the long-wave infrared (LWIR) is critical for applications from autonomous navigation to medical screening, yet existing uncooled detectors are fundamentally constrained. Resistive bolometers are limited by electronic noise and the parasitic thermal load of wired readouts, while state-of-the-art nanomechanical resonators typically rely on vacuum packaging to maintain the mechanical $Q$ needed for sensitivity. Here, we introduce and demonstrate an uncooled thermal detector that addresses these challenges via an all-optical transduction mechanism. The heterogeneously integrated pixel is engineered for minimal thermal mass, combining pyrolytic carbon absorbers for broadband LWIR absorption, hollow zirconia structures for ultra-low-conductance thermal isolation, and a silicon photonic crystal cavity that serves as a high-$Q$ optical thermometer. Operating at ambient temperature and pressure, we measure a specific detectivity of $1.1\times10^{7}$ Jones and a thermal time constant of $27~\mu \mathrm{s}$, corresponding to a speed that surpasses typical high-sensitivity uncooled technologies by an order of magnitude. Based on this detectivity measurement, which is limited by the noise floor of the external optical detection electronics, a physics-based model predicts a $>25\text{-fold}$ performance enhancement to its fundamental thermorefractive-noise-limited value ($3\times10^{8}$ Jones). The optical readout remains functional across ambient and vacuum environments. We expect this architecture to provide a general route toward scalable, high-performance thermal imaging systems.
\end{abstract}

\maketitle
\stoptoc

\section{Introduction} \label{sec:outline}

\begin{figure*}[!htbp]  % You can use [t], [b], [h], or [!htbp] for placement
    \centering
    \includegraphics[width=1\linewidth]{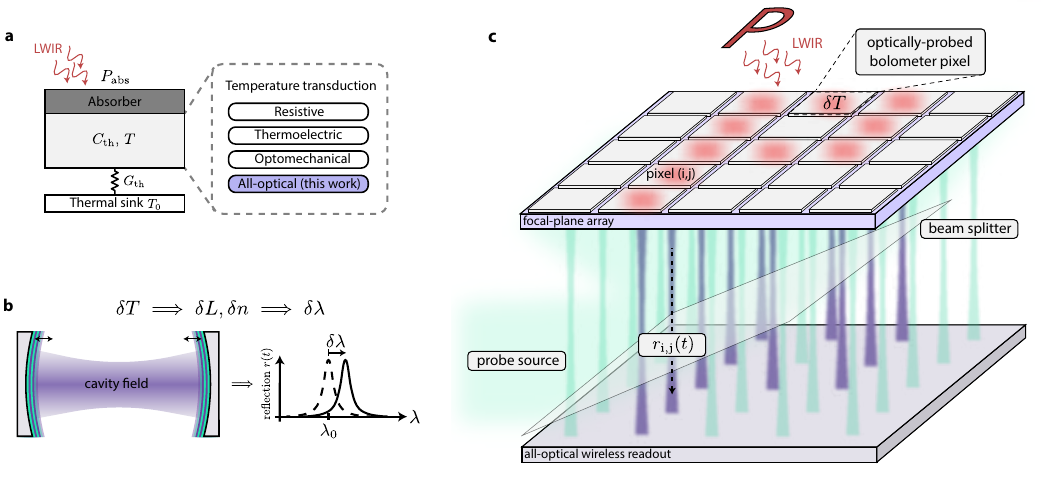} 
    \caption{\textbf{Conceptual framework for an optically-probed bolometer array.} \textbf{a}, A generalized bolometer model. Incident radiation is absorbed ($P_{\text{abs}}$) by a thermal mass ($C_{\text{th}}$) at temperature $T$, which is coupled to a thermal sink ($T_0$) via a thermal conductance ($G_{\text{th}}$). The resulting temperature change ($\delta T$) is transduced by one of several mechanisms. \textbf{b}, Principle of all-optical transduction (this work). The $\delta T$ alters the optical path length of a generic cavity via thermal expansion ($\delta L$) and the thermo-optic effect ($\delta n$), producing a measurable shift ($\delta\lambda$) in its resonance wavelength ($\lambda_0$). \textbf{c}, Conceptual architecture for a wire-free focal-plane array (FPA). A structured LWIR scene (input "P") is imaged onto the FPA, creating a pattern of heated pixels ($\delta T$). A separate probe source (bottom left) illuminates the array via a beam splitter. The probe light (green beams) reflects off the FPA pixels; the reflected light (purple beams) from heated pixels is modulated, carrying the signal $r_{i,j}(t)$, which represents the complex cavity reflection coefficient of the pixel. This reflected signal is directed to the all-optical wireless readout plane (detector array), transducing the thermal image into an optical image.}
    \label{fig:concept}
\end{figure*}

Detection within the thermal infrared atmospheric windows, spanning the mid-wave (MWIR, $3-5~\mathrm{\mu m}$) and long-wave (LWIR, $8-14~\mathrm{\mu m}$), is foundational to modern sensing \cite{rogalski_infrared_2003}. The LWIR band is particularly critical for thermal imaging, as it captures the peak blackbody radiation emitted by objects at room temperature \cite{incropera1990fundamentals}. This principle underpins a growing number of applications, from gas leak detection \cite{zhang_gas_2024} and autonomous driving \cite{rivera_velazquez_analysis_2022, grimming_lwir_2021} to non-invasive medical screening \cite{tattersall_infrared_2016}. As these technologies advance, they place increasingly stringent demands on the speed of the underlying detector hardware, while maintaining high sensitivity.

Cryogenically cooled quantum detectors \cite{rogalski_quantum_2003}, while setting the benchmark for LWIR sensitivity, are hindered from widespread adoption due to the size, weight, power, and cost (SWaP-C) of their cooling systems \cite{rogalski_trends_2021}. This practical limitation has driven the development of uncooled thermal bolometers \cite{yadav_advancements_2022} operating at room temperature.  
The dominant approach, the resistive bolometer \cite{Niklaus, yu_low-cost_2020}, transduces a change in temperature into a change in electrical resistance. This electronic readout, however, is intrinsically limited by Johnson-Nyquist noise and often suffers from low-frequency flicker noise \cite{margaret_kohin_performance_2004, niklaus_uncooled_2007}, in addition to the fundamental speed-sensitivity trade-off common to all thermal detectors. To bypass these electronic noise sources, nanomechanical bolometers \cite{blaikie_fast_2019, PhysRevApplied.9.024016, martini_uncooled_2025, zhang_nanomechanical_2013, jones_mems_2009} have been developed that instead transduce temperature into a mechanical resonance shift. Yet, these advanced platforms have their own practical limitations, including a dependence on vacuum operation for optimal performance \cite{niklaus_uncooled_2007, ekinci_nanoelectromechanical_2005}. 
Combined, these challenges define the key criteria for a next-generation platform: (C1) a path to high detectivity beyond the electronic noise limit; (C2) high-speed (kHz-class) bandwidth for real-time applications; (C3) a pressure-robust transduction mechanism that supports vacuum packaging for fundamental sensitivity limits, without relying on it to enable the readout mechanism; and (C4) scalable, wireless readout architecture to overcome the interconnect density and parasitic thermal load of readout integrated circuits (ROICs). Existing technologies, however, fail to satisfy all four criteria simultaneously: resistive bolometers fundamentally fail (C1-C4), while nanomechanical platforms require vacuum to preserve the mechanical $Q$ needed for sensitivity (C3) and simultaneously maintaining high detectivity and bandwidth remains challenging in practice (C2).

Here, we introduce and demonstrate an uncooled thermal detector that addresses these challenges, circumventing the limitations of both electronic and mechanical platforms by using an all-optical transduction mechanism (see Fig.~\ref{fig:concept}) \cite{PhysRevB.99.205303}. 
The key architectural concept is to minimize the pixel’s heat capacity -- by confining the optical thermometer to a high-$Q$ photonic crystal cavity and coupling it only to low-thermal-mass, electronically passive absorbers -- so that fast response times can be maintained even as the thermal conductance is reduced toward its fundamental limit. This is achieved by suspending the absorber–cavity assembly on zirconia ‘eggshell’ structures, which suppress solid conduction \cite{levi_emerging_2004}.

The photonic crystal (PhC) cavity is fabricated in a commercial silicon photonics foundry \cite{minkov_photonic_2017}. While PhC nanostructures are established platforms for sensing \cite{zhang_review_2015, inan_photonic_2017}, here we use the cavity as an optical thermometer. Incident LWIR radiation is absorbed by heterogeneously integrated pyrolytic carbon nanopillars \cite{mizuno_black_2009, PhysRev.138.A197} and converted to a local temperature rise, which shifts the PhC’s telecom-wavelength resonance via the thermo-optic effect \cite{komma_thermo-optic_2012}. 
This all-optical transduction mechanism (C1, C4) avoids the Johnson–Nyquist and flicker noise sources associated with electronic bolometers \cite{margaret_kohin_performance_2004, niklaus_uncooled_2007}. The complete pixel, assembled using pick-and-place, integrates the cavity, absorbers, and hollow zirconia supports in a compact footprint. Operating under ambient temperature and pressure (C3), the device achieves a specific detectivity of $D^* \approx 1.1 \times 10^7~\mathrm{Jones}$ with a thermal time constant of $\tau_{\text{th}}=27~\mu\mathrm{s}$ (C2), defined in the linear, small-signal transient regime (Supplementary Sec. S4), an order-of-magnitude improvement over typical high-sensitivity uncooled technologies. Inverse-designed optimization of the PhC cavity is expected to reduce the readout to the thermorefractive-noise limit, yielding $D^* \approx 3 \times 10^8~\mathrm{Jones}$ for the present thermal design.

\section{Results} \label{sec:develop}

\subsection{Device architecture and operating principle}

\begin{figure*}[!htbp]  % You can use [t], [b], [h], or [!htbp] for placement
    \centering
    \includegraphics[width=0.98\linewidth]{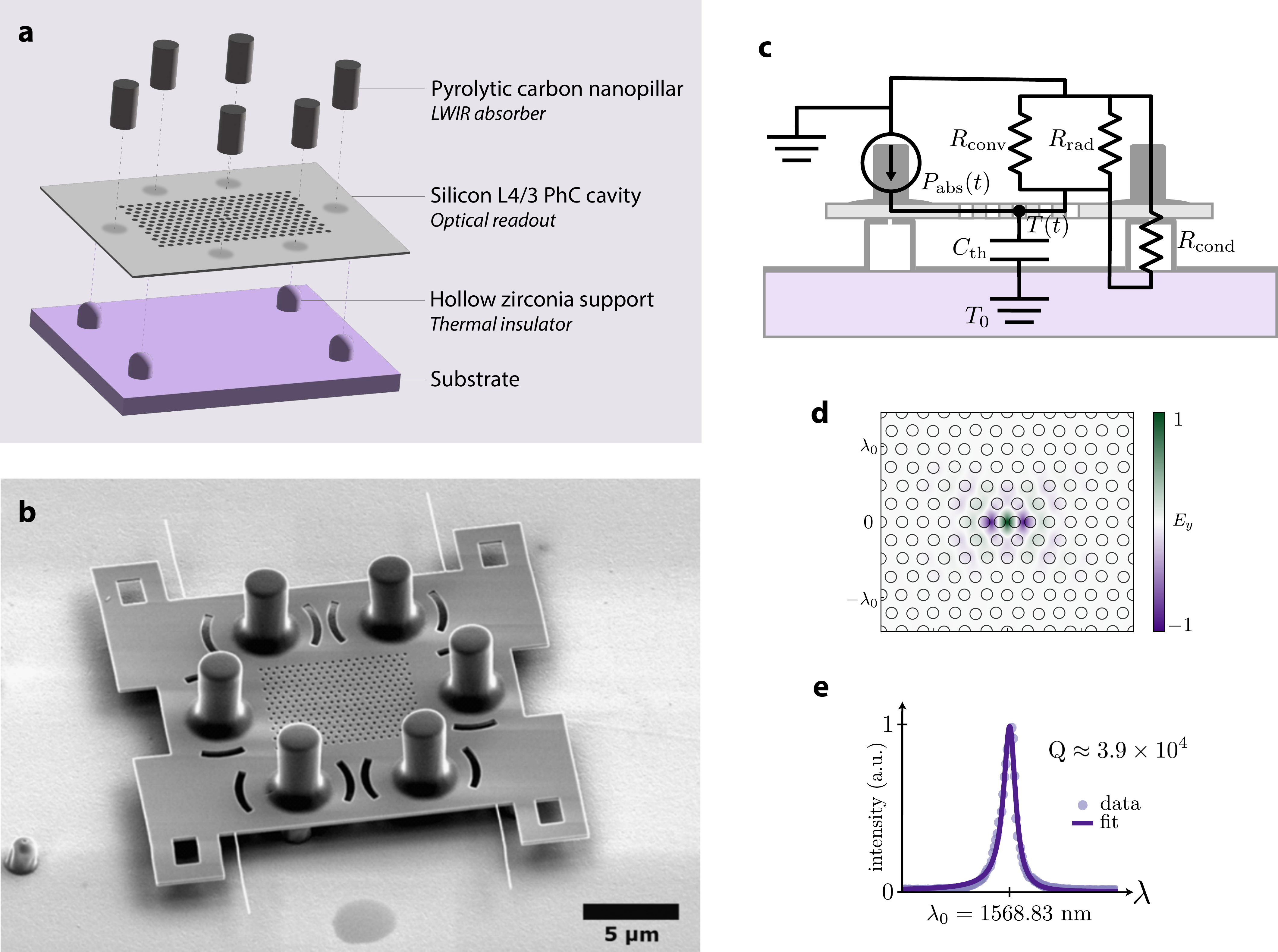} 
    \caption{\textbf{Device architecture and operating principle of the photonic crystal bolometer.} \textbf{a}, 3D schematic of the heterogeneously integrated pixel. The device consists of an array of pyrolytic carbon nanopillars that function as LWIR absorbers, a silicon L4/3 PhC cavity for the optical readout, and hollow zirconia support that provide thermal insulation from the substrate. \textbf{b}, Scanning electron micrograph of a fabricated device. Scale bar, $5~\mu \mathrm{m}$. \textbf{c}, Lumped-element thermal model of the bolometer pixel. We model the device as a thermal RC circuit where the suspended pixel has a total heat capacity $C_{\text{th}}$ and is connected to a thermal bath $(T_0)$ through parallel thermal paths for conduction $(R_{\text{cond}})$, convection $(R_{\text{conv}})$, and radiation $(R_{\text{rad}})$. \textbf{d}, Simulated (finite-difference time-domain) mode profile of a L4/3 PhC cavity. \textbf{e}, Measured optical resonance on the final device, centered at $1568.83~\mathrm{nm}$ and showing a quality factor of $3.9 \times 10^4$.}
    \label{fig:device}
\end{figure*}

Figure~\ref{fig:device} presents the architecture and operating principle of our photonic crystal bolometer. A schematic and scanning electron micrograph of the fabricated pixel are shown in Fig.~\ref{fig:device}a and Fig.~\ref{fig:device}b, respectively. The device is a heterogeneously integrated system designed to transduce LWIR radiation into an optical signal at room temperature and over tunable gas pressures.
The operating principle relies on detecting the spectral shift, $\delta \lambda$, of a high-$Q$ optical microresonator in response to a temperature change, $\delta T$ (Fig.~\ref{fig:concept}b), induced by absorbed LWIR power, $\delta P_{\text{abs}}$. This relationship is quantified by the spectral responsivity \cite{richards_bolometers_1994}:
\begin{equation}
    \mathcal{R_{\lambda}}(\omega) = \dfrac{\delta \lambda}{\delta P_{\text{abs}}}
    = \lambda_0 \left( \alpha + \dfrac{1}{n} \dfrac{dn}{dT} \right) \dfrac{1}{G_{\text{th}}+i \omega C_{\text{th}}}
\label{responsivity}
\end{equation}
where $\lambda_0$ is the resonance wavelength, $\alpha$ and $\dfrac{dn}{dT}$ are the material's thermal expansion and thermo-optic coefficients \cite{komma_thermo-optic_2012, swenson_recommended_1983}, and $G_{\text{th}}$ and $C_{\text{th}}$ are the pixel's total thermal conductance and heat capacity, respectively. 

Since the device's characteristic length $(\sim 20~ \mu \mathrm{m})$ is much larger than the phonon mean free path in silicon at room temperature $(\sim 300~\mathrm{nm})$ \cite{ju_phonon_1999, PhysRevB.101.115301, lee_graphene-based_2020}, the heat flow is macroscopic and the entire pixel can be treated as a single thermal node. 
We then describe its behavior by a lumped-element thermal model equivalent to a first-order RC circuit (Fig.~\ref{fig:device}c) \cite{incropera1990fundamentals, blaikie_fast_2019, richards_bolometers_1994}. 
In this model, the total thermal conductance to the environment, $G_{\text{th}} = 1/R_{\text{th}}$, is the sum of parallel contributions from solid-state conduction $(G_{\text{cond}})$ through the support pillars to the substrate, convection $(G_{\text{conv}})$ to the surrounding air, and thermal radiation $(G_{\text{rad}})$, treated linearly for the small temperature changes induced in our experiment (see Supplementary Section S4).

Achieving high responsivity therefore requires an architecture that minimizes both $G_{\text{th}}$ and $C_{\text{th}}$, while a high quality factor increases the transduction gain \cite{PhysRevB.99.205303}, enabling a more precise measurement of the resonance wavelength against the readout's technical noise floor.

We engineered our device to optimize these parameters through the integration of three key components. First, for the optical readout, we employ a 220 nm-thick silicon L4/3 PhC cavity \cite{minkov_photonic_2017}, fabricated by a commercial foundry (Applied Nanotools Inc.). It provides a narrow resonance at telecom wavelengths ($\lambda_0 \approx 1569~\mathrm{nm}$) with a tightly confined optical mode (Fig.~\ref{fig:device}d) and a measured $Q$ of $\approx 3.9 \times 10^4$ (Fig.~\ref{fig:device}e), while having an ultra-low heat capacity (see Supplementary Section~S5) for an optical resonator. 
Second, to provide broadband absorption while maintaining a minimal total thermal mass, we fabricate pyrolytic carbon nanopillars that provide high absorptance (up to $0.8$) across the LWIR band (Supplementary Figure~S2). These components result in a total estimated pixel heat capacity of $C_{\text{th}} \approx 3.04 \times 10^{-10}~\mathrm{J}/\mathrm{K}$ (see Supplementary Section~S5).
Finally, to achieve strong thermal isolation (low $G_{\text{th}}$), we engineer hollow zirconia "eggshell" supports (see Supplementary Figure~S3). The nanoscale shell thickness ($\approx 20\text{--}40$ nm) constrains phonon transport, thereby suppressing thermal conductivity below bulk limits and minimizing the heat path to the substrate. 
These components are assembled into the final pixel architecture using tungsten-probe pick-and-place.

\subsection{Photothermal responsivity}
\label{subsec: photothermal responsivity}

\begin{figure*}[!htbp]  % You can use [t], [b], [h], or [!htbp] for placement
    \centering
    \includegraphics[width=1\linewidth]{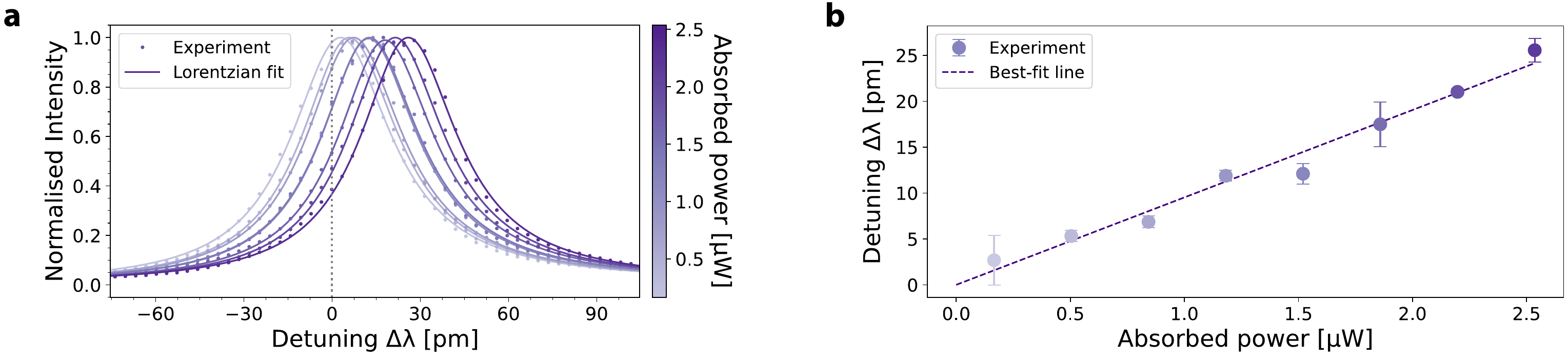}  
    \caption{\textbf{Photothermal responsivity characterization.} \textbf{a}, Measured reflection spectra of the PhC cavity resonance for increasing absorbed optical power, from $0$ to $2.5~\mu \mathrm{W}$. The experimental data (dots) are fitted to Lorentzian lineshapes (solid lines), revealing a systematic red-shift with applied power. \textbf{b}, Extracted resonance detuning $(\Delta \lambda)$ as a function of absorbed power. The linear fit (dashed line) confirms the expected thermo-optic response, and its slope yields the DC photothermal responsivity of the device.}
    \label{fig:DC responsivity}
\end{figure*}

To experimentally quantify the bolometer's photothermal responsivity, we used a $405~\mathrm{nm}$ laser as a controllable, localized heat source. The laser was focused onto a single pyrolytic carbon nanopillar while the optical power incident on the device was varied, and a high-resolution reflection spectrum of the PhC cavity was recorded for each power level (see Methods and Supplementary Figure~S5). The resulting spectra, shown in Fig.~\ref{fig:DC responsivity}a, exhibit a systematic red-shift with increasing power. This behavior is consistent with a device in a thermal steady-state, where the temperature rise is linearly proportional to the absorbed power via the pixel's thermal resistance, $R_{th} =1/G_{th}$. The temperature increase induces this wavelength shift primarily through the thermo-optic effect; for silicon, this contribution is more than an order of magnitude larger than that of thermal expansion and is therefore the dominant transduction mechanism (see Supplementary Table~S1) \cite{komma_thermo-optic_2012, swenson_recommended_1983}.

To analyze this response quantitatively, we plot the resonance detuning, $\Delta \lambda$, as a function of absorbed power in Fig.~\ref{fig:DC responsivity}b. The absorbed power, $P_{\text{abs}}$, is calculated from the measured incident power using a literature-based absorptance of $\eta_{\mathrm{abs}} \approx 91\%$ \cite{1985umo..rept.....Q} for pyrolytic carbon at $405~\mathrm{nm}$. The data confirm a highly linear relationship, from which we extract the DC photothermal responsivity, $\mathcal{R}_{\lambda, DC}$. From the slope of the linear fit, we determine a responsivity of $\mathcal{R}_{\lambda, DC}=9.52 \times 10^3~\mathrm{nm}/\mathrm{W}$.

This responsivity measurement allows for a direct calculation of the pixel's effective thermal resistance. Rearranging Eq.~\ref{responsivity} for the steady-state case and using the known material properties of silicon, we extract an effective thermal resistance for the pixel of $R_{\text{th}} = 1/G_{\text{th}} \approx 1.08 \times 10^5~\mathrm{K}/\mathrm{W}$. This effective key parameter governs not only the static responsivity but also the temporal response of the device.

\subsection{Thermal dynamics and bandwidth}

The response speed of the bolometer is governed by the heat transport dynamics within the pixel. As predicted by our thermal circuit model, in the small-signal regime (Supplementary Sec. S4), the device's response to a step-change in optical power will follow an exponential transient governed by a single thermal time constant, $\tau_{\text{th}} = R_{\text{th}} C_{\text{th}}$.

\begin{figure*}[!htbp]  % You can use [t], [b], [h], or [!htbp] for placement
    \centering
    \includegraphics[width=1\linewidth]{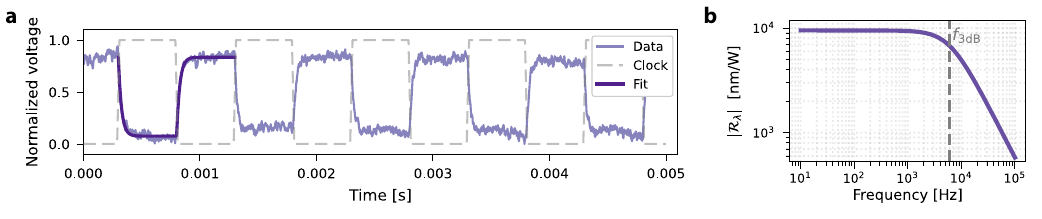}  
    \caption{\textbf{Thermal dynamics and bandwidth of the PhC bolometer.} \textbf{a}, Measured time-domain response to a $1~\mathrm{kHz}$ square-wave modulated heating laser. The homodyne voltage (purple trace), which tracks the pixel's temperature, is fitted (dark purple trace) to our linearized lumped-element, single-pole thermal RC model, yielding a time constant of $\tau_{\text{th}} = 27~\mu \mathrm{s}$. \textbf{b}, Calculated frequency-dependent responsivity. It follows a first-order low-pass filter response, with a $3~\mathrm{dB}$ bandwidth of $f_{\mathrm{3dB}} \approx 5.9~\mathrm{kHz}$ determined from the measured time constant.}
    \label{fig:speed}
\end{figure*}

To measure this time constant experimentally, we use a balanced homodyne interferometer (see Supplementary Section~S3) \cite{yuen_noise_1983}.
The homodyne interferometer converts the deterministic shift of the PhC cavity resonance into a measurable voltage, $v_h(t)$ \cite{PhysRevX.10.041046, PhysRevB.99.205303}. 
We modulated the $405~\mathrm{nm}$ heating laser with a $1~\mathrm{kHz}$ square wave and recorded the temporal response of the homodyne voltage, which directly tracks the resonance wavelength (Fig.~\ref{fig:speed}a). The $1~\mathrm{kHz}$ modulation frequency was chosen to be within the device's operational bandwidth, ensuring a high-fidelity measurement of its transient response. By fitting to our linearized, single-pole thermal RC model, we extract an experimental thermal time constant of $\tau_{\text{th}} = 27~\mu \mathrm{s}$. 

The measured value provides a direct consistency check for our thermal model. Using the thermal resistance determined in Sec.~\ref{subsec: photothermal responsivity} $(R_{\text{th}} \approx 1.08 \times 10^5~\mathrm{K}/\mathrm{W})$ and the total heat capacity estimated as $C_\mathrm{th} \approx 3.04 \times 10^{-10}~\mathrm{J/K}$ (see Supplementary Section~S5), we independently predict a time constant of $\tau_{\text{th}} = R_{\text{th}} C_{\text{th}} \approx 33~\mu \mathrm{s}$. The close agreement between measured and predicted values confirm the self-consistency of our model and the accuracy of the extracted thermal parameters.

This measured time constant corresponds to a $3\mathrm{dB}$ bandwidth of $f_{3\mathrm{dB}} = 1 / (2 \pi \tau_{\text{th}}) \approx 5.9~\mathrm{kHz}$, a value that enables applications in high-speed thermal analysis \cite{grimming_lwir_2021, rivera_velazquez_analysis_2022}. The full frequency-dependent responsivity, calculated using the low-pass filter model, is shown in Fig.\ref{fig:speed}b.

\subsection{Optical readout noise and sensitivity}

To determine the bolometer's sensitivity, we first characterized the noise floor of our optical readout. We achieve this using the same homodyne interferometer setup which converts the PhC cavity's phase fluctuations, $\delta \phi_{cav}(t)$, into the voltage $v_h(t)$ \cite{PhysRevX.10.041046, PhysRevB.99.205303}. Near resonance, these phase fluctuations are linearly proportional to the wavelength fluctuations, $\delta \lambda(t)$, allowing us to relate the voltage power spectral density (PSD), $S_v(f)$, to the wavelength PSD, $S_{\lambda}(f)$:

\[
S_{\lambda}(f) = \frac{S_v(f)}{S^2}, \quad \text{where } S = \left. \frac{dv_h}{d\lambda} \right|_{\lambda_{\text{op}}}
\]

\begin{figure*}[!htbp]  % You can use [t], [b], [h], or [!htbp] for placement
    \centering
    \includegraphics[width=1\linewidth]{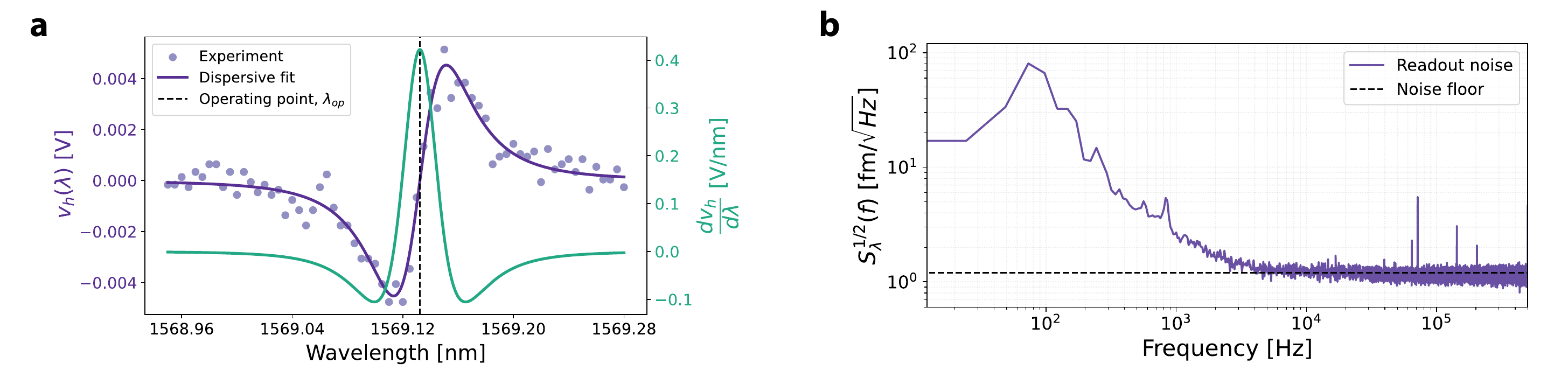}  
    \caption{\textbf{Optical readout calibration and noise characterization.} \textbf{a}, Calibration of the wavelength-to-voltage transduction. The homodyne voltage signal $(v_h)$ is measured as the probe laser is swept across the cavity resonance (dots), showing excellent agreement with a fit to a dispersive theoretical model (solid purple line). The right axis shows the calculated transduction gain $(dv_h/d \lambda)$. The vertical line marks the optimal operating point $(\lambda_{\text{op}})$ of maximum gain. \textbf{b}, Wavelength noise amplitude spectral density of the readout. The spectrum is characterized by technical noise at low frequencies and a white noise plateau at higher frequencies. The dashed line indicates the measured instrument-limited noise floor of $1.2~\mathrm{fm}/\sqrt{\mathrm{Hz}}$.}
    \label{fig:Noise}
\end{figure*}

To experimentally determine the calibration factor, $S$, we measured the dispersive lineshape of the homodyne signal by sweeping the probe laser across the resonance (Fig.~\ref{fig:Noise}a). The data are in good agreement with a fit to the derivative of a Lorentzian, the expected theoretical lineshape for a phase-sensitive measurement at quadrature (see Supplementary Section~S6.2). For the noise measurement, a PID controller actively locks the interferometer to the quadrature point of maximum slope, $\lambda_{\text{op}}$, ensuring the highest transduction gain. 

A systematic noise budget analysis was performed to identify the limiting noise source in our setup (see Supplementary Section~S6.1). Operation in the shot-noise-limited regime, which requires high LO power, was found to introduce spurious interferometric effects from optical back-reflections, which compromised the integrity of the calibration measurement. To ensure an accurate calibration, we therefore performed the noise characterization at a moderate LO power where the system was limited by the detector's electronic noise floor. The final calibrated wavelength noise PSD, shown in Fig.~\ref{fig:Noise}b, was obtained by recording a continuous time-domain voltage trace, converting it to a PSD using Welch's method \cite{welch1967use}, and applying the calibration factor $S$. The spectrum reveals low-frequency technical noise before reaching a white noise plateau, from which we measure a readout noise floor of $\sqrt{S_{\lambda}} = 1.2~\mathrm{fm}/\sqrt{\mathrm{Hz}}$. This value represents the instrument-limited sensitivity of our readout and is not yet limited by the fundamental thermodynamic performance of the resonator, which is set by the thermo-refractive noise (TRN) fluctuations \cite{PhysRevX.10.041046}. 

\subsection{Frequency-dependent specific detectivity}

The specific detectivity $(D^*)$ \cite{nudelman_detectivity_1962} is a key figure of merit that normalizes the bolometer's sensitivity to its physical area, allowing for direct comparison between different detector technologies. We calculate it by combining the results from the preceding sections using the relation:
\begin{equation}
    D^* = \dfrac{\sqrt{A_d} |\mathcal{R}_{\lambda}(f)|}{\sqrt{S_{\lambda}(f)}}
\end{equation}
where $A_d \approx 4 \times 10^{-6}~\mathrm{cm}^2$ is the physical area of the pixel, $|\mathcal{R}_{\lambda}(f)|$ is the frequency-dependent responsivity, and $\sqrt{S_{\lambda}(f)}$ is the wavelength noise amplitude spectral density.

\begin{figure}[!htbp]  % You can use [t], [b], [h], or [!htbp] for placement
    \centering
    \includegraphics[width=\columnwidth]{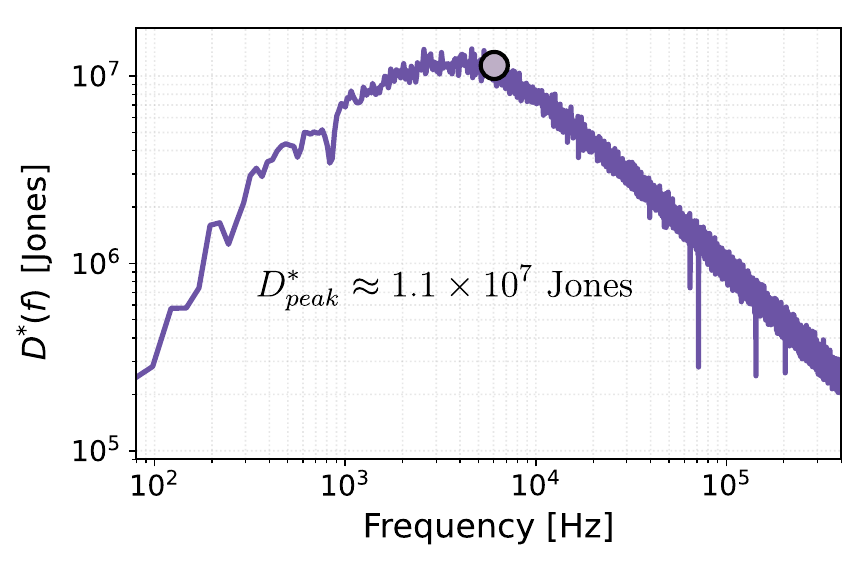}  
    \caption{\textbf{Frequency-dependent specific detectivity.} The calculated $D^*$ spectrum for the bolometer, showing a peak value of $1.1 \times 10^7~\mathrm{Jones}$ near $6~\mathrm{kHz}$ (black circle). The performance is limited by technical noise below $1~\mathrm{kHz}$, while the high-frequency roll-off is determined by the thermal time constant.}
    \label{fig:Detectivity}
\end{figure}

The resulting $D^*$ spectrum, plotted in Fig.~\ref{fig:Detectivity}, reveals a peak performance of $D^* \approx 1.1 \times 10^7~\mathrm{Jones}$ (where $1~\text{Jones} = 1~\text{cm}\cdot\text{Hz}^{1/2}\cdot\text{W}^{-1}$) around $5~\mathrm{kHz}$, near the device's thermal cutoff frequency. The shape of the spectrum is governed by the previously characterized device parameters. At low frequencies, $D^*$ is suppressed by the technical noise of the readout system, while at frequencies above the peak, it rolls off as a direct consequence of the thermal time constant limiting the responsivity.    

\section{Discussions} \label{sec:develop}

\begin{figure*}[!htbp]  % You can use [t], [b], [h], or [!htbp] for placement
    \centering
    \includegraphics[width=0.8\linewidth]{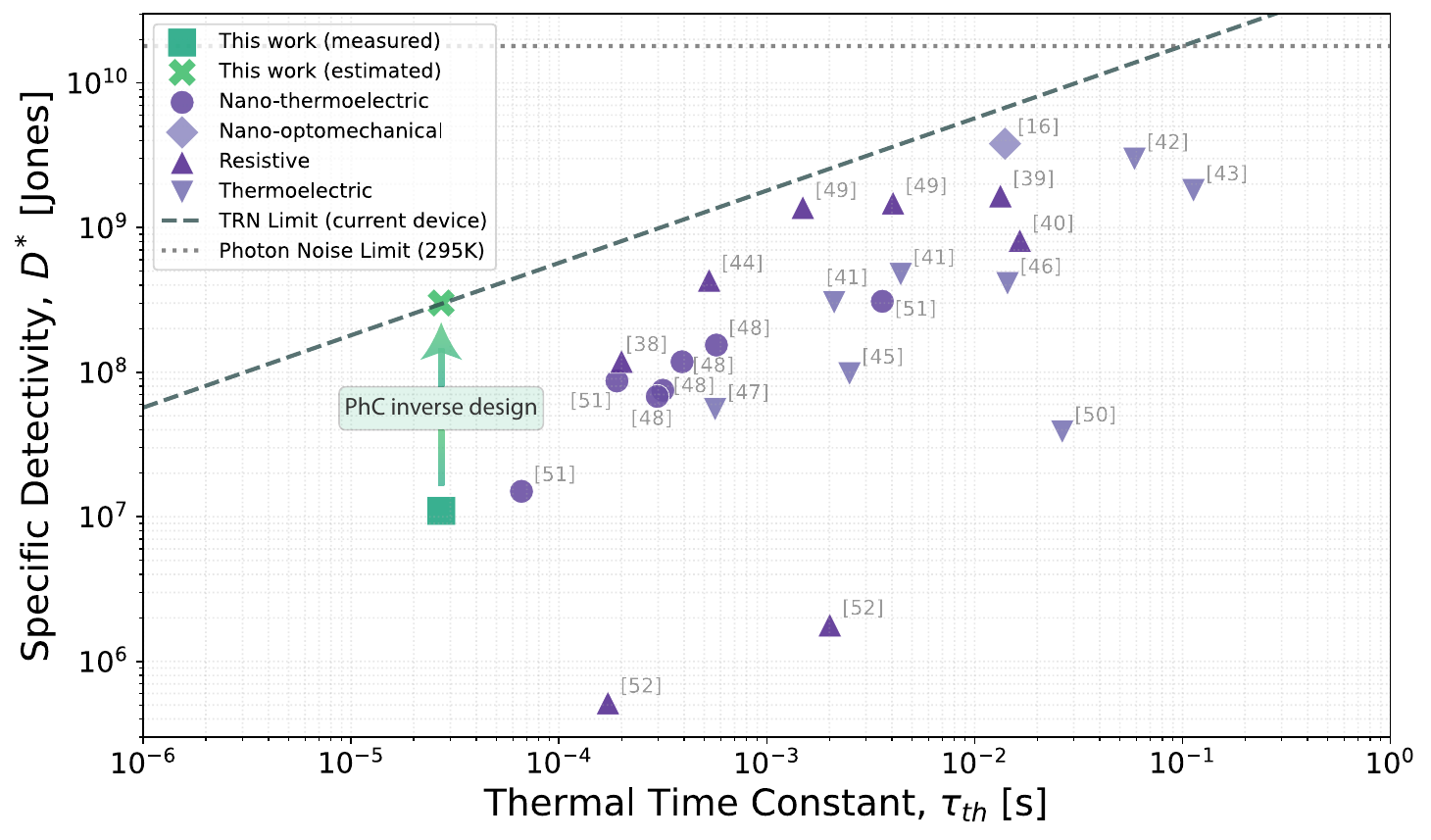} 
    \caption{\textbf{Performance benchmark of uncooled LWIR detectors.} Specific detectivity ($D^*$) as a function of thermal time constant $(\tau_{\text{th}})$ for various uncooled LWIR detector technologies \cite{almasri_self-supporting_2001, dong_uncooled_2002, dong_uncooled_2003, foote2003thermopile, gawarikar_high_2013, haenschke_new_2012, karanth_infrared_2009, ke_design_2018, lei_double-end-beam_2016, li_front-side_2019, martini_uncooled_2025, murros_infrared_2025, schnelle1984bolometers, shen_performance_2019, varpula_nano-thermoelectric_2021, zhou_improvements_2018}. This work (green square) is benchmarked against representative resistive, thermoelectric and nano-optomechanical bolometers from the literature. The green cross marks the projected performance of the same device with inverse-designed PhC cavity optimization, corresponding to the calculated thermo-refractive-noise (TRN) limit ($D^*_{\mathrm{limit}}\approx3\times10^{8}~\mathrm{Jones}$), indicated by the dashed green line for the present thermal design (with pixel heat capacity $C_{\text{th}}$). The dotted grey line denotes the photon noise limit for a 295~K background.}
    %\textbf{(Add diode and transistor based bolometers?)}}
    \label{fig:benchmark}
\end{figure*}

% Summary and Benchmarking
Our photonic crystal bolometer achieves a specific detectivity of $D^* \approx 1.1 \times 10^7~\mathrm{Jones}$ and a thermal time constant of $27~\mu \mathrm{s}$ at ambient temperature and pressure. This performance point, shown in Fig.~\ref{fig:benchmark}, is distinguished by its speed, as the $5.9~\mathrm{kHz}$ bandwidth is more than an order of magnitude faster than that of many high-sensitivity uncooled bolometers, including resistive and nano-thermoelectric platforms.

For a single-mode bolometer limited by TRN \cite{PhysRevX.10.041046}, the fundamental detectivity is set by the thermal conductance, $D^*_\mathrm{limit} \propto 1/\sqrt{G_\mathrm{th}}$, or equivalently $D^*_\mathrm{limit} \propto \sqrt{\tau_{\mathrm{th}}/C_\mathrm{th}}$ at fixed $C_\mathrm{th}$ in the linear regime mentioned earlier. Thus, for a given pixel heat capacity, the TRN-limited performance follows a single $D^*_\mathrm{limit}(\tau_{\mathrm{th}})$ curve along which changes in $G_\mathrm{th}$ move the operating point. 
For the present pixel, based on the measured thermal resistance and calculated heat capacity, this trade-off is shown by a dashed green line in Fig.~\ref{fig:benchmark}, and the green cross marks the corresponding TRN-limited operating point $D^*_\mathrm{limit} \approx 3 \times 10^8~\mathrm{Jones}$. 
Lowering $G_\mathrm{th}$ -- for example by operating in vacuum to suppress convection -- would increase $D^*_\mathrm{limit}$ at the expense of a longer thermal time constant. Because our architecture enables a small pixel heat capacity, the corresponding TRN curve in Fig.~\ref{fig:benchmark} lies above the envelope of reported uncooled LWIR detectors over the $\tau_{\mathrm{th}}$ range shown, so that at a given thermal time constant a TRN-limited implementation of this architecture could, in principle, achieve higher specific detectivity.

The key innovation enabling this performance is the use of a photonic resonator for the readout, distinguishing it from both non-resonant electric transducers and resonant mechanical sensors. This approach provides several advantages. First, it is inherently immune to the Johnson-Nyquist noise of resistive elements (C1). Second, its low heat capacity allows for high-speed operation (C2). Third, its wire-free optical readout provides a scalable FPA architecture, avoiding the interconnect density and parasitic thermal load of ROICs (C4). Other resonant bolometers based on nanomechanical systems require vacuum operation to prevent viscous air damping from degrading the mechanical quality factor \cite{ekinci_nanoelectromechanical_2005, LandauLifshitz1982FluidMechanics}, a fundamental requirement for their readout. Our optical cavity readout, by contrast, is insensitive to the surrounding pressure (C3), which can be convenient for operation at ambient conditions.

% Significance of the Approach
The performance of our device demonstrates the viability of a heterogeneously integrated photonic platform for thermal sensing. While our system's sensitivity is currently limited by technical noise in our off-chip measurement apparatus, the bolometer pixel itself is not subject to the on-chip electronic noise sources that can limit conventional devices. The architecture, which combines foundry-fabricated silicon photonics with specialized, separately-engineered components, provides a flexible foundation for future development.
% Analysis of Current Limitations
% Pathways to Improvement
We identify three distinct pathways for future improvement toward the fundamental limits of our platform: enhancing the optical readout, optimizing the bolometer's thermal properties, and developing scalable manufacturing techniques. 

First, the system's detectivity is currently limited by the electronic noise of our external apparatus, $(1.2~\mathrm{fm}/\sqrt{\mathrm{Hz}})$, not by the fundamental (C1) thermodynamics of the sensor itself. The ultimate performance of the current design is set by its TRN floor \cite{PhysRevX.10.041046}, which we calculate to be $\sqrt{S_{\lambda, TRN}} \approx 0.064~\mathrm{fm}/\sqrt{\mathrm{Hz}}$ (see Supplementary Section~S6.3). This corresponds to our estimated $D^*_\mathrm{limit}$. Bridging this gap requires improving the signal-to-noise ratio of the readout. Since the calibrated wavelength noise is given by $S_{\lambda} = S_v/S^2$, with a transduction gain $S \propto \eta^{1/2}Q$ (see Supplementary Section~S7), performance can be enhanced by increasing both the cavity's zero-order diffraction efficiency $(\eta)$ and quality factor. Such cavities, with an optimized far-field mode (near-unity $\eta$) and a $Q$-factor exceeding $10^6$, have been demonstrated in a previous work \cite{panuski_full_2022}, using inverse design \cite{minkov_inverse_2020}.

Second, improvements beyond the TRN limit require re-engineering the pixel’s thermal properties \cite{PhysRevX.10.041046}. Further reductions in $C_\mathrm{th}$ through pixel miniaturization or elimination of residual parasitic thermal mass would shift this entire curve upward, increasing the achievable detectivity at any given $\tau_\mathrm{th}$. Beyond these linear optimizations, the platform also supports operation in the nonlinear thermal regime (see Supplementary Section S4.2), where temperature-dependent heat loss channels (e.g., radiative $\propto T^4$) render $G_\text{th}$ effectively state dependent. This can yield strongly non-exponential transients and enable amplitude-dependent response dynamics that are not captured by a single fixed $RC$ constant, offering a rich parameter space for future high-speed transient sensing.

Finally, to translate this single-pixel performance to practical imaging systems, the proof-of-concept tungsten-probe assembly must be transitioned to a high-throughput, scalable manufacturing process, such as micro-transfer printing with elastomeric stamps \cite{kaur_hybrid_2021}, successfully realized for similar devices (C4) \cite{bommer_transfer_2025}.

% Conclusion and Outlook
In conclusion, we have demonstrated a new class of uncooled thermal detector that leverages a heterogeneously integrated, optically-probed photonic crystal cavity to deliver kilohertz-bandwidth response while maintaining high detectivity. 
The device attains a specific detectivity of $D^* \approx 1.1\times10^{7}~\mathrm{Jones}$ together with a fast thermal time constant of $\tau_{\mathrm{th}}=27~\mu\mathrm{s}$ (C2), representing a significant advance demonstrated under ambient-pressure conditions.
By eliminating the on-chip electronic readout, our architecture is not fundamentally limited by Johnson-Nyquist noise (C1) and is amenable to scalable manufacturing through advanced integration techniques (C4). 
Further analysis indicates a potential $>25\times$ improvement beyond our present instrument-limited performance toward the thermo-refractive-noise limit of $D^*_{\mathrm{limit}} \approx 3\times10^{8}~\mathrm{Jones}$. This frontier, set by the pixel’s small heat capacity, places our current thermal design on a performance envelope that exceeds reported uncooled LWIR detectors. Notably, the optical readout remains functional across ambient and vacuum environments (C3).
This positions our platform as a strong candidate for the next generation of high-speed, large-format focal-plane arrays.
Operating without the need for cryogenic cooling, such arrays could enable a wide range of applications, from real-time industrial process monitoring \cite{zhang_gas_2024} and autonomous navigation \cite{rivera_velazquez_analysis_2022, grimming_lwir_2021} to advanced scientific imaging \cite{tattersall_infrared_2016}.

\section{Methods} \label{sec:method}

\subsection{Device Fabrication}

The bolometer pixel was constructed from three separately fabricated components: a silicon PhC cavity, pyrolytic carbon absorbers, and hollow zirconia supports. A detailed schematic of the fabrication and assembly process is provided in Supplementary Figure~S1. The PhC cavity was procured from a commercial foundry (Applied Nanotools Inc.), where it was patterned into the top layer of a silicon-on-insulator (SOI) wafer using positive-tone e-beam lithography and reactive ion etching (RIE). The SOI wafer consisted of a $220~\mathrm{nm}$-thick silicon device layer on a 2,000 nm buried oxide (BOX) layer. The cavity design is an L4/3 line-defect in a triangular lattice of air holes, with a periodic grating perturbation applied to the positions of adjacent holes to enhance vertical out-coupling efficiency \cite{minkov_photonic_2017}.

The pyrolytic carbon absorbers were fabricated directly onto the PhC device die. The surface was coated with a diluted mr-DWL 40 negative photoresist, which was patterned via photolithography to form polymer pillars with pre-pyrolysis diameters of $4-6~\mu \mathrm{m}$. These pillars were pyrolyzed by ramping the temperature to $800\,^{\circ}\mathrm{C}$ in a \ce{H2}:\ce{Ar} ambient environment. After pyrolysis, which yielded final pillar diameters of $3-4~\mu \mathrm{m}$, a gentle oxygen plasma was applied to the device to remove carbonaceous residues from the fabrication process. Concurrently, the hollow zirconia supports were fabricated on a separate transparent glass substrate. Sacrificial polymer molds were patterned from a diluted mr-DWL 5 photoresist, coated with a $20-40~\mathrm{nm}$ thick layer of $\mathrm{ZrO_{2}}$ via atomic layer deposition (ALD), and drilled with a $\sim 100~\mathrm{nm}$ hole using a focused ion beam (FIB). The polymer core was then removed through the hole with an oxygen plasma asher, leaving the hollow supports.

For the final assembly, the PhC and absorber structure was first released from its handle wafer by a hydrofluoric (HF) acid etch that selectively removes the BOX layer, leaving a suspended "stamp". Using a micromanipulator-based system, this stamp was then broken free with a tungsten tip and transferred onto the hollow zirconia supports on the final glass substrate.

\subsection{Optical Characterization}

\textbf{Experimental Apparatus.} The device was characterized at ambient conditions using a custom-built, cross-polarized free-space microscope detailed in Supplementary Fig.~S5. The probe beam was sourced from a tunable C-band telecom laser (Santec TSL-710), while a 405 nm diode laser provided a photothermal stimulus. The two beams were combined via a dichroic mirror and focused onto the device through a single 0.55 NA objective. A set of steering mirrors in the 405 nm laser path allowed its focused spot to be positioned onto a specific pyrolytic carbon pillar, independent of the probe beam, which was centered on the PhC cavity. The input probe polarization was set at 45° relative to the cavity's principal axis using a half-wave plate, enabling the polarization-rotated cavity reflection to be isolated from specular reflections via a polarizing beamsplitter. The sample was mounted on a stage temperature-stabilized to within 10 mK by a Peltier plate with a feedback controller. For phase-sensitive measurements, a balanced homodyne detection path was used, where the reflected probe signal was interfered with a path-length-matched local oscillator (LO) and detected by a balanced photodetector (Thorlabs PDB480C).\\

\textbf{Characterization Procedures.} To measure the photothermal responsivity, we used a direct reflection path to record the cavity spectrum on an IR camera (Xenics Cheetah 640). The 405 nm laser power was varied in a randomized sequence, with each measurement repeated three times for statistical robustness. The resonance wavelength was extracted from each spectrum by fitting to a Lorentzian lineshape. For noise and temporal characterization, we utilized the homodyne detection path. Measurements were performed at a moderate LO power where the system was limited by the detector's electronic noise floor to ensure an accurate calibration. The transduction gain was determined by performing an open-loop sweep of the probe laser's wavelength across the resonance. For measurements, a PID controller actively locked the interferometer to the quadrature point of maximum slope using a piezo-actuated mirror in the LO path. The temporal response was measured by modulating the 405 nm laser with a square wave and recording the homodyne voltage on a digital oscilloscope.
The noise PSD was then calculated from the recorded time-domain traces using Welch's method \cite{welch1967use}.\\

\section{Acknowledgments}
The authors acknowledge Mahmoud Jalali Mehrabad, Chao Li and Saumil Bandyopadhyay for valuable discussions. The authors thank Flexcompute for supporting finite-difference time-domain
simulation. This material is based upon work sponsored by the U.S. Army DEVCOM ARL Army Research Office through the MIT Institute for Soldier Nanotechnologies under Cooperative Agreement number W911NF-23-2-0121.\\

\section{Data and code availability}
The data and code supporting the findings of this study are available at: https://github.com/FolletLouis/PhC-bolometer.\\

\section{COMPETING INTERESTS}
J.G., C.L.P. and D.R.E are inventors on U.S. Patent No. 11,635,330 B2, which relates to the technology described in this work. Other authors declare no competing interests. \\

\section{Author contributions}
J.G., C.L.P, and D.R.E. conceived the idea. L.F. performed the experiments, data analysis, simulations and modeling. I.C. and S.T.-M. supported the experiments. J.G. and C.L.P. designed and fabricated the device and conducted initial characterization. L.F. wrote the manuscript. All authors discussed the results and revised the manuscript. D.R.E. supervised the work.\\

\bibliography{bib}

\clearpage

\onecolumngrid  % SI is commonly single-column

% ---- S-numbering ----
% \setcounter{page}{1}       \renewcommand{\thepage}{\arabic{page}}
\setcounter{section}{0}    \renewcommand{\thesection}{S\arabic{section}}
\setcounter{subsection}{0} \renewcommand{\thesubsection}{S\arabic{section}.\arabic{subsection}}
\setcounter{figure}{0}     \renewcommand{\thefigure}{S\arabic{figure}}
\setcounter{table}{0}      \renewcommand{\thetable}{S\arabic{table}}
\setcounter{equation}{0}   \renewcommand{\theequation}{S\arabic{equation}}

% ---- Make sure subsection refs don't prepend the section again ----
\makeatletter
\renewcommand{\p@subsection}{}% no extra prefix for \ref of a subsection
% (good hygiene with hyperref anchors)
\renewcommand{\theHsection}{S\arabic{section}}
\renewcommand{\theHsubsection}{S\arabic{section}.\arabic{subsection}}
\makeatother

\begin{center}
    \vspace*{1em}
    {\large \textbf{Supplementary Information:}}\\[0.5em]
    {\large \textbf{All-Optical Photonic Crystal Bolometer with Ultra-Low Heat Capacity for Scalable Thermal Imaging}}\\[1em]

    Louis Follet$^{1}$,
    Jordan Goldstein$^{1}$,
    Christopher L.~Panuski$^{1}$,
    Ian Christen$^{1}$,
    Sivan Trajtenberg-Mills$^{2}$,
    and Dirk R.~Englund$^{1}$\\[0.7em]

    {\small\itshape
    $^{1}$Research Laboratory of Electronics, Massachusetts Institute of Technology, Cambridge, MA 02139, USA\\
    % $^{2}$Electrical Engineering and Computer Science, Massachusetts Institute of Technology, Cambridge, MA 02139, USA\\
    $^{2}$School of Electrical Engineering, Faculty of Engineering, Tel Aviv University, Tel Aviv 6997801, Israel
    }\\[0.7em]

\end{center}

\bigskip

\resumetoc
\tableofcontents

\newpage

\section{Device Fabrication}
\label{sec:device fab}

\begin{figure*}[!htbp]  % You can use [t], [b], [h], or [!htbp] for placement
    \centering
    \includegraphics[width=1\linewidth]{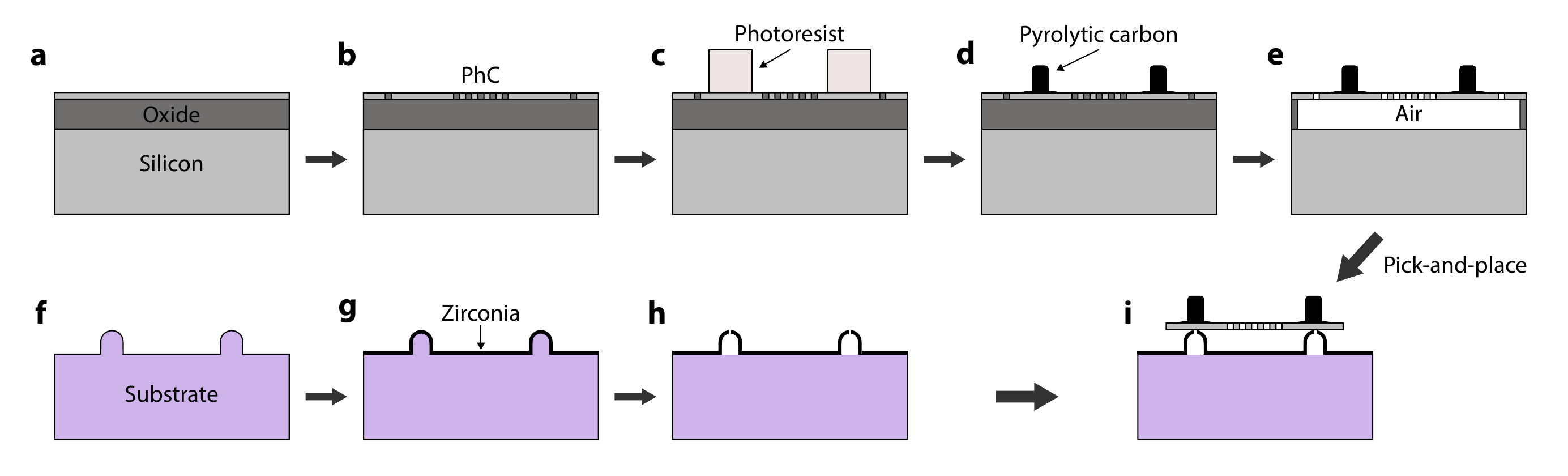} 
    \caption{\textbf{Fabrication and heterogeneous integration of the photonic crystal bolometer.} \textbf{(a-e)} Fabrication of the suspended device "stamp". \textbf{(a)} The process begins with a SOI wafer, which consists of a silicon device layer on top of a buried oxide layer. \textbf{(b)} A PhC cavity is patterned into the silicon device layer. \textbf{(c)} Photoresist pillars are defined on the PhC die, and \textbf{(d)} are subsequently converted into pyrolytic carbon absorbers through a high-temperature pyrolysis process. \textbf{(e)} The underlying buried oxide layer is selectively removed with a vapor HF etch to release the suspended stamp. \textbf{(f-h)} Concurrent fabrication of the hollow supports. \textbf{(f)} On a separate substrate, sacrificial polymer molds are patterned. \textbf{(g)} A conformal layer of zirconia is deposited via ALD. \textbf{(h)} The polymer core is removed using an $\mathrm{O_{2}}$ plasma ash, leaving hollow supports. \textbf{(i)} In the final step, the released PhC stamp is transferred via a pick-and-place technique onto the zirconia supports to form the final device.}
    \label{fig:SI_fab}
\end{figure*}

The bolometer pixel is assembled from three separately fabricated components: (i) a silicon photonic-crystal (PhC) cavity patterned in a 220 nm device layer of an SOI wafer, (ii) an array of pyrolytic-carbon nanopillars that provide broadband LWIR absorption, and (iii) hollow zirconia supports that thermally isolate the suspended pixel from the substrate.
A schematic and sequence of the process are shown in Fig~\ref{fig:SI_fab}.

\subsection{Silicon PhC cavity}
\label{subsec:device fab - phc}

PhC cavities were fabricated by a commercial foundry (Applied Nanotools Inc.) in the 220 nm silicon device layer of an SOI wafer with a 2 µm buried oxide (BOX).
The design is an L4/3 line-defect cavity with a weak periodic grating perturbation that enhances out-of-plane coupling \cite{minkov_photonic_2017}.
Pattern transfer used positive-tone e-beam lithography and reactive-ion etching (RIE).

\subsection{Pyrolytic-carbon absorbers}
\label{subsec:device fab - absorbers}

The PhC die was coated with a diluted mr-DWL 40 negative resist, exposed to define polymer pillars of $4-6~\mu \mathrm{m}$ diameter, and pyrolyzed at $800\,^{\circ}\mathrm{C}$ in \ce{H2}:\ce{Ar} \cite{mizuno_black_2009}.
After pyrolysis, the pillars shrink to $3-4~\mu \mathrm{m}$ diameter and retain high optical absorption with minimal mass; a brief $\mathrm{O_{2}}$ plasma removes surface residue.

\begin{figure*}[!htbp]  % You can use [t], [b], [h], or [!htbp] for placement
    \centering
    \includegraphics[width=0.4\linewidth]{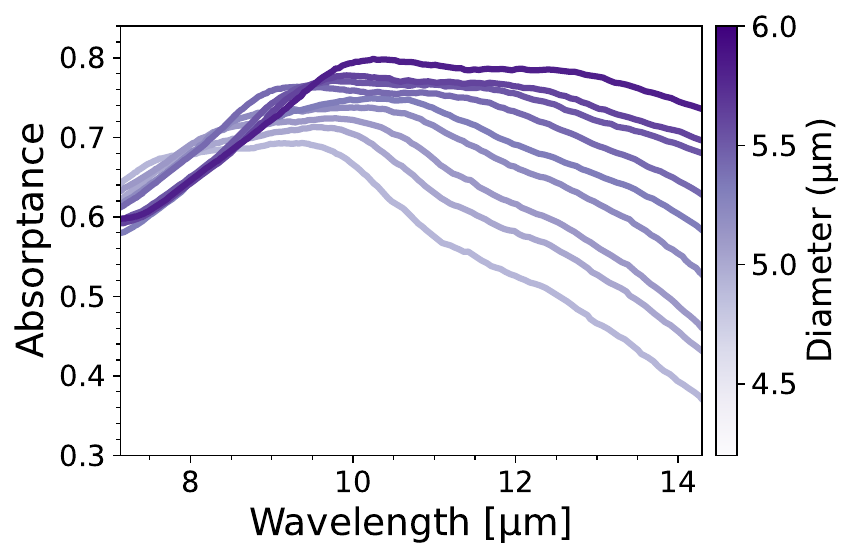} 
    \caption{\textbf{FTIR absorptance of pyrolytic-carbon pillars.} Absorptance vs wavelength for pillar arrays of different final diameters. The data confirm broadband, high absorptance in the LWIR range.}
    \label{fig:FTIR}
\end{figure*}

FTIR of the fabricated pyrolytic-carbon pillars (See Fig~\ref{fig:FTIR}.) shows broad absorptance across the LWIR ($8-14~\mu \mathrm{m}$) range, peaking around 0.8 near $10~\mu \mathrm{m}$, along with dependence on pillar diameter.

\subsection{Hollow zirconia supports}
\label{subsec:device fab - eggshells}

On a separate transparent substrate, sacrificial polymer nubbins (mr-DWL 5) were defined and conformally coated by atomic-layer deposition (ALD) of $\mathrm{ZrO_{2}}$ ($20-40~\mathrm{nm}$) to form closed shells \cite{levi_emerging_2004}.

\begin{figure*}[!htbp]  % You can use [t], [b], [h], or [!htbp] for placement
    \centering
    \includegraphics[width=0.9\linewidth]{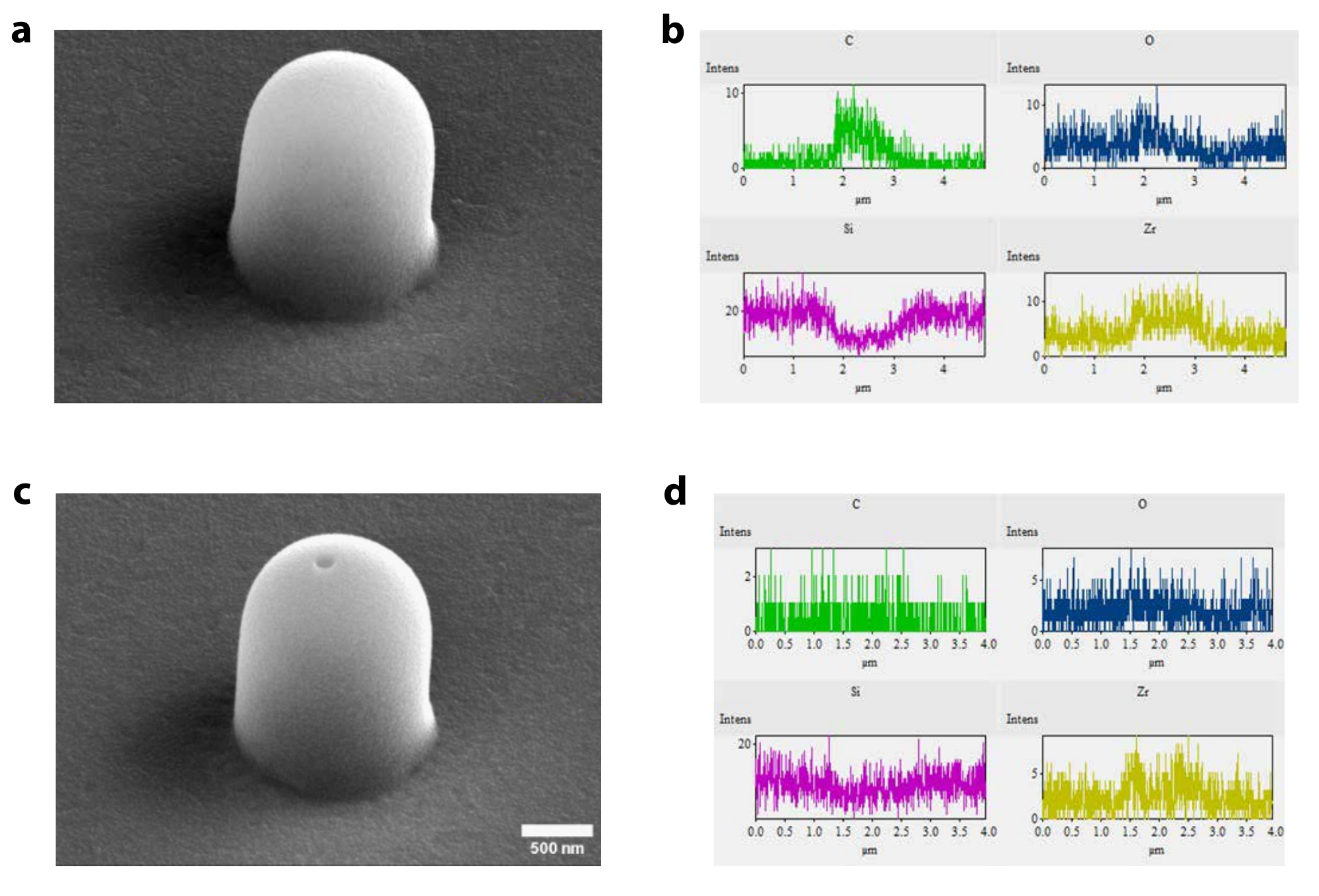} 
    \caption{\textbf{Role of the FIB vent in hollowing zirconia “eggshells.”} \textbf{a,b}, SEM and EDS line-scans of a $\mathrm{ZrO_{2}}$-coated polymer nubbin before venting: a pronounced carbon peak indicates the retained polymer core.
    (c,d) After FIB milling a small vent at the shell apex, the internal carbon signal vanishes while zirconium and oxygen remain, confirming that the vent allows the polymer to be fully removed by the oxygen plasma ashing and leaves a hollow $\mathrm{ZrO_{2}}$ shell.}
    \label{fig:eggshell_FIB}
\end{figure*}

A small $\approx 100~\mathrm{nm}$ focused-ion-beam (FIB) vent was opened in the shell, and the internal polymer was removed by $\mathrm{O_2}$ plasma ashing. 
Energy-dispersive X-ray spectroscopy (EDS) line-scans acquired in the SEM highlight the role of the FIB vent in the hollowing process (see Fig~\ref{fig:eggshell_FIB}.). Without the vent, the polymer core remains largely trapped beneath the $\mathrm{ZrO_{2}}$ coating after $\mathrm{O_2}$ plasma ashing, as indicated by a strong internal carbon signal. After introducing the FIB hole and performing the same plasma ashing, the carbon signal drops to background while zirconium and oxygen remain, confirming that the vent enables complete removal of the organic core and yields a hollow ceramic shell.

\subsection{Integration}
\label{subsec:device fab - integration}

\begin{figure*}[!htbp]  % You can use [t], [b], [h], or [!htbp] for placement
    \centering
    \includegraphics[width=0.5\linewidth]{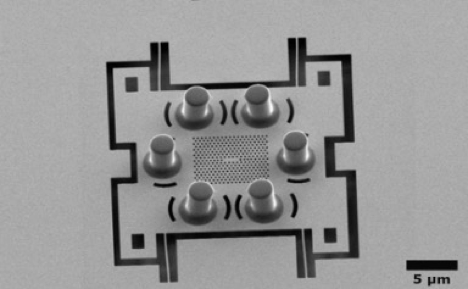} 
    \caption{\textbf{Released silicon “stamp” prior to transfer.”} SEM of the suspended silicon membrane carrying the L4/3 photonic-crystal cavity and six surrounding pyrolytic-carbon absorbers. Narrow silicon tethers ($\approx 100-200~\mathrm{nm}$) retain the membrane on the frame after BOX removal and are gently fractured during pick-and-place.}
    \label{fig:stamp}
\end{figure*}

For the final assembly, the PhC and absorber structure was first released from its handle wafer by a hydrofluoric (HF) acid etch that selectively removed the buried oxide (BOX) layer, leaving a suspended silicon “stamp” connected to the surrounding frame by narrow silicon tethers ($\approx 100-200~\mathrm{nm}$ wide), as shown in Fig~\ref{fig:stamp}.
These tethers mechanically support the membrane after release and allow it to be manipulated under a microscope without damage.

Using a micromanipulator-based tungsten probe, the suspended stamp was then broken free at the tether anchors and transferred onto the pre-fabricated hollow $\mathrm{ZrO_{2}}$ supports on the target glass substrate, as illustrated in Fig~\ref{fig:SI_fab}i. This step completes the heterogeneous pixel.

\section{Experimental Setup}
\label{sec:setup}

\begin{figure*}[!htbp]  % You can use [t], [b], [h], or [!htbp] for placement
    \centering
    \includegraphics[width=0.9\linewidth]{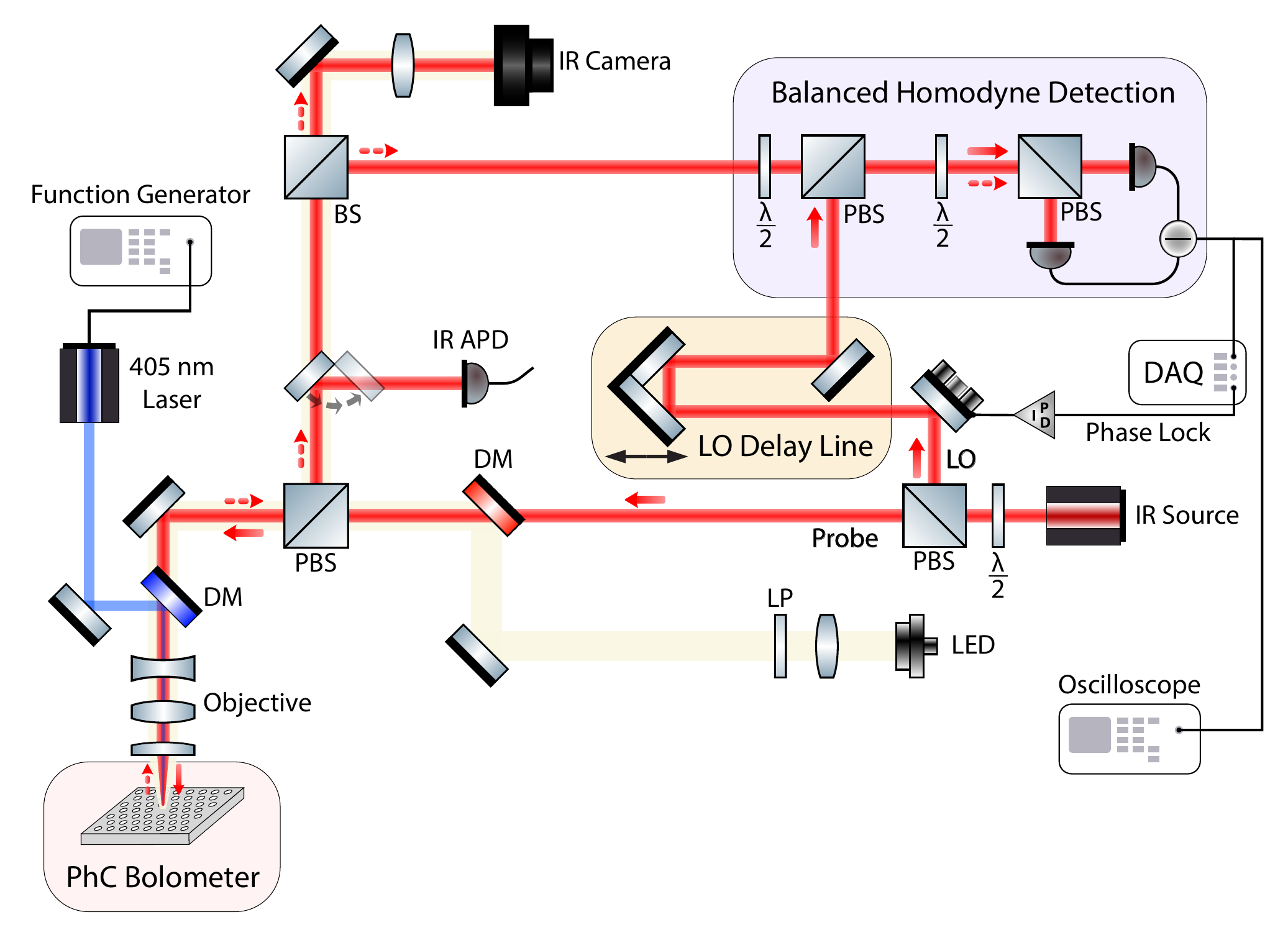} 
    \caption{\textbf{Optical characterization setup.} Schematic of the custom-built, cross-polarized free-space microscope used for all optical characterization. A tunable C-band telecom laser (IR Source) provides the probe beam, which is directed to the PhC bolometer through a polarizing beamsplitter (PBS) and objective. A 405 nm diode laser, driven by a function generator for modulation, provides the photothermal stimulus. The reflected probe signal (dashed red arrow) is separated from the input path by the PBS. For photothermal responsivity measurements, the signal is directed by a beamsplitter (BS) to an IR Camera to record the cavity spectrum. For phase-sensitive noise and temporal response measurements, the signal is sent to the Balanced Homodyne Detection module. In this path, the signal is interfered with a path-length-matched local oscillator (LO) and detected by a balanced photodiode pair. For noise floor characterization, the 405 nm laser is off. The homodyne voltage is recorded on a digital Oscilloscope. An additional IR APD path is available for direct power measurements, and a white light LED path is used for alignment. BS: 90:10 Beamsplitter; PBS: Polarizing Beamsplitter; DM: Dichroic Mirror; LP: Linear Polarizer; $\lambda / 2$: Half-wave plate; LO: Local Oscillator; DAQ: Data Acquisition system.}
    \label{fig:SI_setup}
\end{figure*}

All optical measurements were performed using a custom-built, cross-polarized free-space microscope shown in Fig~\ref{fig:SI_setup}, and detailed in Methods in the main text.
The tunable C-band telecom laser provided the probe beam, which was focused onto the PhC bolometer through a polarizing beamsplitter and high-numerical-aperture (NA = 0.55) objective.
A 405 nm diode laser, intensity-modulated by a function generator, served as the photothermal excitation source.
The reflected probe was routed to either an IR camera for cavity spectra or a balanced homodyne detector for phase-sensitive measurements of noise and temporal response.
The setup also included an auxiliary InGaAs APD for power monitoring and a white-light LED path for device alignment and focusing.

\section{Balanced homodyne interferometry}
\label{sec:BHI}

\subsection{Principle}
\label{subsec:BHI - principle}

Homodyne detection measures a chosen quadrature of an optical field by interfering the field under test (our ``signal'') with a strong, phase-coherent local oscillator (LO) at the same optical frequency, and reading out the intensity difference between two complementary output ports of a $50{:}50$ beamsplitter. The difference signal is linear in the signal field and (because the two ports carry the same mean LO power) exhibits strong common-mode rejection of LO intensity noise \cite{yuen_noise_1983}. 

The realization is shown in Supplementary Fig~\ref{fig:SI_setup}. The probe reflected from the PhC cavity provides the signal; the LO is derived from the same laser and routed through a phase/delay line. The two fields are brought into a common spatial mode with orthogonal linear polarizations by a polarizing beamsplitter (PBS). A $\lambda/2$ plate sets the analysis basis to $\pm 45^{\circ}$, and a second PBS projects the mixed field onto two outputs that are detected on a balanced photodiode pair. In this polarization-domain analyzer, the $\lambda/2{+}$PBS block implements the $50{:}50$ mixing of signal and LO required for homodyne detection, and the subsequent electronic subtraction yields the balanced output.

\subsection{Field relations and balanced difference}
\label{subsec:BHI - field relations}

Let the fields incident on the analysis PBS (after the $\lambda/2$ plate) be
\begin{equation}
\mathbf{E}_{\mathrm{sig}}(t) = \sqrt{P_{\mathrm{sig}}}\,\chi(\lambda)\, e^{i \omega t}\, \hat{\mathbf{e}}_x, 
\qquad
\mathbf{E}_{\mathrm{LO}}(t) = \sqrt{P_{\mathrm{LO}}}\, e^{i(\omega t + \phi_{\mathrm{LO}})}\, \hat{\mathbf{e}}_y,
\end{equation}
with $\hat{\mathbf{e}}_{x,y}$ orthogonal linear polarizations, $P_{\mathrm{sig}}$ the collected signal power, and $\chi(\lambda)$ the complex field response of the cavity in reflection. Define 
\[
\phi \equiv \phi_{\mathrm{LO}} + \arg \chi.
\]
Rotating to the $\pm45^{\circ}$ basis and splitting on the analysis PBS yields output fields
\begin{equation}
\mathbf{E}_{\pm} = \frac{1}{\sqrt{2}}\left(\mathbf{E}_{\mathrm{LO}} \pm \mathbf{E}_{\mathrm{sig}}\right),
\end{equation}
so that the detected powers are
\begin{equation}
P_{\pm} = \frac{P_{\mathrm{LO}} + P_{\mathrm{sig}}|\chi|^2}{2} 
    \pm \sqrt{P_{\mathrm{LO}} P_{\mathrm{sig}}}\, 
    \mathrm{Re}\!\left[\chi(\lambda)e^{i\phi}\right].
\end{equation}

With photodiode responsivity $\mathcal{R}$ (A/W), the balanced photocurrent difference is
\begin{equation}
i_{\Delta} \equiv \mathcal{R}\,(P_{+} - P_{-})
    = 2 \mathcal{R} \sqrt{P_{\mathrm{LO}} P_{\mathrm{sig}}}\,
    \mathrm{Re}\!\left[\chi(\lambda)e^{i\phi}\right]
    \;\propto\; X_{\phi},
\end{equation}
where $X_{\phi}$ denotes the signal quadrature, and the quadrature angle $\phi$ is set by the LO phase. Locking the interferometer with the active PID controller such that $\phi = \pi/2$ selects the phase quadrature, which is the operating condition used for our measurements, as it provides the maximum linear slope of the cavity response and minimizes sensitivity to residual amplitude noise.

\section{Thermal regimes and time constants}
\label{sec:thermal regimes}

We model the pixel temperature field $T(\mathbf r,t)$ (silicon membrane + PhC cavity, pyrolytic-carbon pillars, zirconia supports) by
\begin{equation}
  \rho_j c_{p,j}\,\frac{\partial T_j}{\partial t}
  = \nabla\!\cdot\!\bigl(k_j \nabla T_j\bigr) + Q_j(\mathbf r,t),
  \qquad j \in \{\text{Si, pc, ZrO}_2\},
  \label{eq:heat_pde_full}
\end{equation}
with continuity of $T$ and normal heat flux at internal interfaces. On surfaces exposed to the gas, we impose
\begin{equation}
  -k_s \frac{\partial T}{\partial n}
  = h\bigl(T,T_\infty\bigr)\,\bigl(T - T_\infty\bigr)
    + \varepsilon\sigma\bigl(T^4 - T_\infty^4\bigr)
  \quad\text{on } S_{\text{gas}},
  \label{eq:bc_gas_full}
\end{equation}
and at the supports into the substrate
\begin{equation}
  -k_{\text{ZrO}_2}\frac{\partial T}{\partial n}
  = G_{\text{cond}}^{(\text{eff})}\,\bigl(T - T_0\bigr)
  \quad\text{on } S_{\text{sub}}.
  \label{eq:bc_sub_full}
\end{equation}
Equations~\eqref{eq:heat_pde_full}–\eqref{eq:bc_sub_full} define the first-principles thermal model. The measured output is the resonance shift of the PhC cavity,
\begin{equation}
  \delta\lambda(t) = \Lambda_T \,\bar{\theta}(t), \qquad
  \bar{\theta}(t) = \frac{1}{V_{\text{pix}}}
  \int_{\Omega_{\text{pix}}} \bigl[T(\mathbf r,t)-T_0\bigr]\,dV,
  \label{eq:output_def}
\end{equation}
with
\begin{equation}
  \Lambda_T = \lambda_0\!\left(\alpha + \frac{1}{n}\frac{dn}{dT}\right)
\end{equation}
the thermo-optic/expansion coupling for silicon.
The various operating regimes can be organized as follows.

\subsection{Equilibrium (quasi-static) regime}

If the absorbed power $P_{\text{abs}}(t)$ varies slowly compared to the intrinsic thermal relaxation times of the pixel, the time derivative in Eq.~\eqref{eq:heat_pde_full} may be neglected and the device is effectively in steady state:
\begin{equation}
  \nabla\!\cdot\!\bigl(k \nabla T_{\text{eq}}\bigr)
  + Q\bigl(\mathbf r; P_{\text{abs}}\bigr) = 0
\end{equation}
with boundary conditions~\eqref{eq:bc_gas_full}–\eqref{eq:bc_sub_full}. The pixel-averaged temperature rise
$\bar{\theta}_{\text{eq}}(P_{\text{abs}})$ defines the static (DC) responsivity via
\begin{equation}
  R_{\lambda,\text{DC}} \equiv \frac{\delta\lambda}{\delta P_{\text{abs}}}
  = \Lambda_T \,\frac{d\bar{\theta}_{\text{eq}}}{dP_{\text{abs}}}.
\end{equation}
No explicit time constant appears at this level: the measured rise/fall of the readout is controlled by the temporal variation of the scene or heating source rather than by the detector’s intrinsic dynamics.

\subsection{Transient regime: general case}

When $P_{\text{abs}}(t)$ changes on time scales comparable to or faster than the detector response, the full time-dependent problem~\eqref{eq:heat_pde_full} must be solved. It is convenient to write $T(\mathbf r,t) = T_0 + \theta(\mathbf r,t)$ with $\theta$ the temperature rise.

We distinguish two cases in the transient regime:\\

\paragraph{Linear small-signal approximation.}

For sufficiently small excursions $|\theta|\ll T_0$, material and boundary parameters can be linearized around
$T_0$:
\begin{equation}
  k(T) \approx k_0,\quad
  c_p(T) \approx c_{p,0},\quad
  h\bigl(T,T_\infty\bigr) \approx h_0,
\end{equation}
and
\begin{equation}
  T^4 - T_0^4 \approx 4 T_0^3\,\theta.
\end{equation}
The governing equation becomes
\begin{equation}
  \rho c_p\,\frac{\partial \theta}{\partial t}
  = \nabla\!\cdot\!\bigl(k \nabla \theta\bigr) + q(\mathbf r,t),
  \label{eq:heat_lin}
\end{equation}
with linear Robin boundary conditions,
\begin{equation}
  -k\frac{\partial \theta}{\partial n}
  = G_{\text{surf}}\,\theta \quad\text{on }S_{\text{gas}},\qquad
  -k\frac{\partial \theta}{\partial n}
  = G_{\text{cond}}^{(\text{eff})}\,\theta \quad\text{on }S_{\text{sub}},
\end{equation}
where $G_{\text{surf}} = h_0 + 4\varepsilon\sigma T_0^3$. This defines a linear, time-invariant diffusion operator
$\mathcal L$,
\begin{equation}
  \frac{\partial \theta}{\partial t} = \mathcal L \theta + s(\mathbf r,t).
  \label{eq: operator}
\end{equation}
The homogeneous problem admits a modal expansion
\begin{equation}
  \theta(\mathbf r,t) = \sum_n a_n(t)\,\phi_n(\mathbf r),
  \qquad
  \mathcal L\phi_n = -\frac{1}{\tau_n}\phi_n,
\end{equation}
with modal amplitudes
\begin{equation}
  a_n(t) =
  \begin{cases}
    A_n\bigl(1-e^{-t/\tau_n}\bigr), & \text{step heating},\\[2pt]
    B_n e^{-t/\tau_n}, & \text{step cooling}.
  \end{cases}
  \label{eq:modal amp}
\end{equation}
The set of time constants $\{\tau_n\}$ is fixed by $\mathcal L$ and is therefore the same for heating and cooling; rise and decay differ only in the coefficients $A_n,B_n$. The impulse response is a sum of decaying exponentials with these $\tau_n$.

If internal diffusion within the pixel is much faster than the leakage to the bath, the slowest eigenmode is nearly uniform and higher-order modes decay quickly. In that case the dynamics of the pixel-averaged temperature $\bar{\theta}(t)$ may be approximated by a single degree of freedom with total heat capacity
\begin{equation}
  C_{\text{th}} = \sum_j \rho_j c_{p,j}V_j
\end{equation}
and total thermal conductance $G_{\text{th}}$:
\begin{equation}
  C_{\text{th}}\frac{d\bar{\theta}}{dt}
  + G_{\text{th}}\,\bar{\theta}
  = P_{\text{abs}}(t).
  \label{eq:lumped_ode}
\end{equation}
This is the standard one-pole (RC) model with a single thermal time constant
\begin{equation}
  \tau_{\text{th}} = \frac{C_{\text{th}}}{G_{\text{th}}}.
\end{equation}
For a step in $P_{\text{abs}}$ the rise and decay of $\bar{\theta}(t)$ are then strictly single-exponential with the same $\tau_{\text{th}}$.\\

\paragraph{Nonlinear regime (large $\Delta T$ or ultrafast excitation).}

For larger temperature excursions, the approximations leading to Eq.~\eqref{eq:heat_lin} no longer apply:
\begin{itemize}
  \item the radiative term retains its $T^4$ dependence and the effective radiative conductance
        $G_{\text{rad}}(T) = 4\varepsilon\sigma T^3 A$ depends on $T$;
  \item the gas heat-transfer coefficient $h(T,T_\infty)$ can increase with $|T-T_\infty|$;
  \item $k(T)$ and $c_p(T)$ may acquire measurable $T$-dependence.
\end{itemize}
The evolution of the pixel-averaged temperature can then be written schematically as
\begin{equation}
  C_{\text{th}}(T)\,\frac{d\bar{T}}{dt}
  + G_{\text{th}}(T)\,\bigl(\bar{T}-T_0\bigr)
  = P_{\text{abs}}(t),
  \label{eq: NL}
\end{equation}
with $G_{\text{th}}(T)$ and possibly $C_{\text{th}}(T)$ now state-dependent. In this nonlinear regime a single, input-independent time constant is not well-defined. One may introduce a local quantity $\tau_{\text{eff}}(T) = C_{\text{th}}(T)/G_{\text{th}}(T)$, but this varies along the trajectory. Heating and cooling transients are no longer related by a simple sign change, and if separately fitted to a single exponential they can yield different effective “rise” and “decay” time constants. For very short impulses that strongly excite higher thermal modes, the response is multi-exponential already at early times.

\subsection{Illustrative solved transients}

\begin{figure*}[!htbp]  % You can use [t], [b], [h], or [!htbp] for placement
    \centering
    \includegraphics[width=0.9\linewidth]{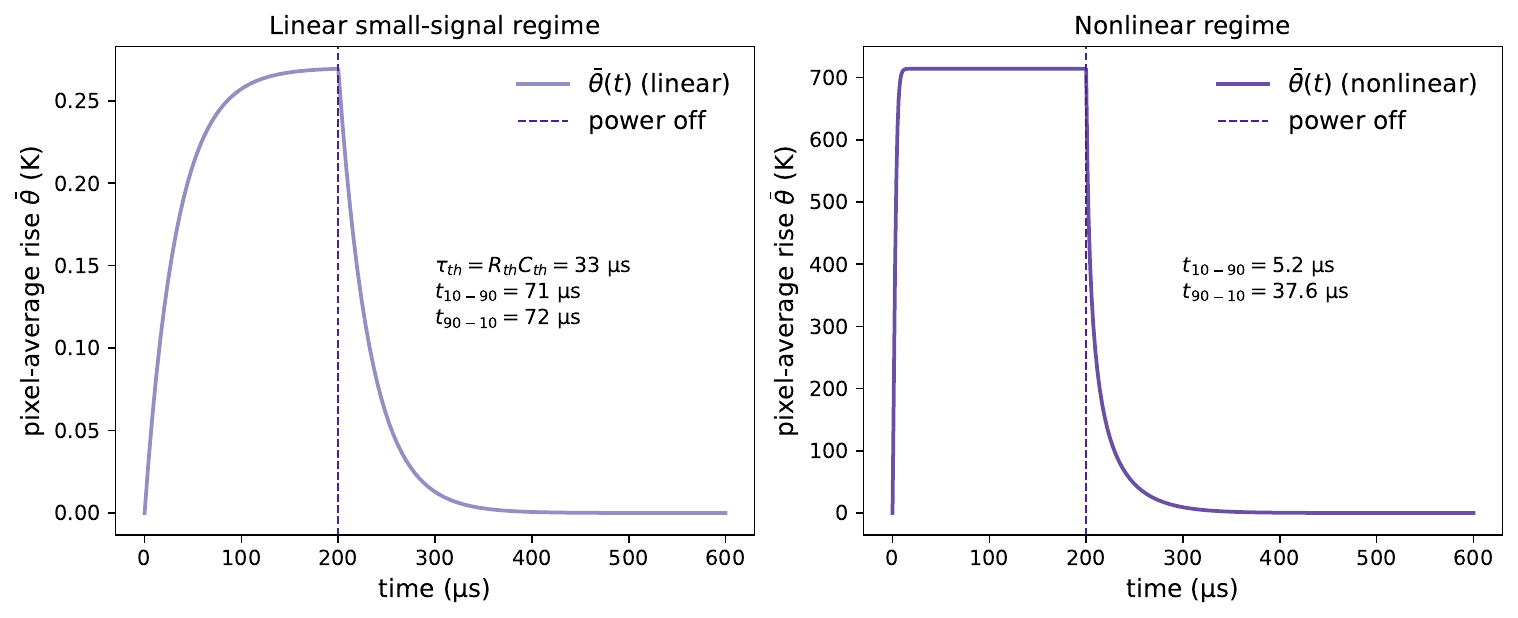} 
    \caption{\textbf{Solved transient response in the linear small‑signal and nonlinear regimes.} Pixel-averaged temperature rise $\bar{\theta}(t)$ in response to a square absorbed-power pulse
$P_{\mathrm{abs}}(t)=P_{\mathrm{on}}$ for $0<t<t_{\mathrm{off}}$ and
$P_{\mathrm{abs}}(t)=0$ for $t\ge t_{\mathrm{off}}$, with
$t_{\mathrm{off}}=200~\mu\mathrm{s}$ (dashed line).
\textbf{Left:} Linear single-mode (one-pole) thermal model with
$R_{\mathrm{th}}=1.08\times10^{5}~\mathrm{K/W}$ and
$C_{\mathrm{th}}=3.04\times10^{-10}~\mathrm{J/K}$,
$P_{\mathrm{on}}=2.5~\mu\mathrm{W}$, and
$\Delta T_{\max}\approx R_{\mathrm{th}}P_{\mathrm{on}}\approx0.27~\mathrm{K}$.
The thermal time constant is
$\tau_{\mathrm{th}}=R_{\mathrm{th}}C_{\mathrm{th}}\approx33~\mu\mathrm{s}$,
yielding identical rise and decay dynamics
($t_{10\text{--}90}\approx t_{90\text{--}10}\approx71~\mu\mathrm{s}$).
\textbf{Right:} Nonlinear illustrative case retaining full radiative loss
(Eqs.~\ref{eq:bc_gas_full}, \ref{eq: NL}) with
$T_{0}=295~\mathrm{K}$,
$\varepsilon=0.8$,
$A_{\mathrm{eff}}=1.0\times10^{-6}~\mathrm{m}^{2}$,
and $P_{\mathrm{on}}=50~\mathrm{mW}$.
The non-radiative conductance is chosen such that the linearization about
$T_{0}$ matches the same small-signal thermal conductance $G_{\mathrm{th}}$, resulting in asymmetric effective rise and decay times
($t_{10\text{--}90}\approx5.2~\mu\mathrm{s}$,
$t_{90\text{--}10}\approx37.6~\mu\mathrm{s}$).
}
    \label{fig:SI_transient_regimes}
\end{figure*}

To visualize the distinction between the linear small-signal transient regime and the nonlinear transient regime discussed above, Fig.~\ref{fig:SI_transient_regimes} plots the pixel-averaged temperature rise $\bar{\theta}(t)$ in response to a square-wave step in absorbed power $P_{\mathrm{abs}}(t)$ that is switched off at $t=t_{\mathrm{off}}$ (vertical dashed line). In the linear regime, linearizing Eqs.~\ref{eq:heat_pde_full}--\ref{eq:bc_sub_full} about $T_0$ yields a linear time-invariant operator $\mathcal{L}$ (Eq.~\ref{eq: operator}), so the set of relaxation times $\{\tau_n\}$ is identical for heating and cooling (Eq.~\ref{eq:modal amp}). In the single-mode limit this reduces to the one-pole model (see Eq.~\ref{eq:lumped_ode}), implying identical characteristic rise and decay times. 

In contrast, retaining the full nonlinear boundary losses (e.g., the $T^4$ radiative term in Eq.~\ref{eq:bc_gas_full} leads to a state-dependent effective conductance $G_{\mathrm{th}}(T)$ (Eq.~\ref{eq: NL}, for which a single, input-independent time constant is not well-defined; heating and cooling transients need not be symmetric.

\subsection{Regime realized in the present measurements}

For the operating conditions reported in the main text ($P_{abs,max} \approx 2.5~\mu$W), the estimated temperature rise is $\Delta T_{max} \approx 0.3$~K $\ll T_{0}$. Furthermore, internal diffusion times in the Si membrane are estimated to be $< 1~\mu$s, significantly faster than the measured response time ($\sim 27~\mu$s).
Consequently, our experiments fall strictly within the \textbf{transient, linear, single-mode regime}. The static responsivity is linear with power (Fig.~\ref{fig:DC responsivity}), and the temporal response (Fig.~\ref{fig:speed}a) is governed by a single, power-independent time constant $\tau_{th}$ determined exclusively by the ratio $C_{th}/G_{th}$.

\section{Heat capacity calculation}
\label{sec:heat capacity}
\newcommand{\Cth}{C_{th}}

Having established the validity of the single-mode approximation in Section~\ref{sec:thermal regimes}, here we compute the pixel's total heat capacity $C_{th}$ from geometry and materials only. We then combine this with the independently measured thermal resistance $R_{th}$ to predict $\tau_{th}^{pred} = R_{th}C_{th}$, which is compared with the directly measured time constant $\tau_{th}^{meas}$.

\subsection{Model and parameters}
\label{subsec:heat capacity - model}

For small temperature perturbations, the device is well described by a single thermal mode \cite{incropera1990fundamentals, richards_bolometers_1994, blaikie_fast_2019} with total heat capacity
\begin{equation}
  C_{th} = \sum_{j} \rho_j \, c_{p,j} \, V_j,
  \qquad
  j \in \{\text{Si membrane}, \ \text{pyrolytic-carbon pillars}\},
\end{equation}
where $\rho$ is mass density, $c_p$ is the (room-temperature) specific heat capacity, and $V$ is volume. The zirconia contribution is neglected (see Sec.~\ref{subsec:heat capacity - assumptions}).

The different numerical values used in following derivations are reported in Supplementary Table \ref{tab:parameters}.\\

\subsection{Component-wise calculation}
\label{subsec:heat capacity - calculation}

\paragraph{Silicon membrane.}
\begin{align}
  V_{\mathrm{Si}}
  &= (23\times10^{-6}~\mathrm{m})(16\times10^{-6}~\mathrm{m})(220\times10^{-9}~\mathrm{m}) \notag\\
  &= 8.096\times10^{-17}~\mathrm{m^{3}}, \\[4pt]
  m_{\mathrm{Si}} &= \rho_{\mathrm{Si}} V_{\mathrm{Si}}
  = 2330~\mathrm{kg\,m^{-3}} \times 8.096\times10^{-17}~\mathrm{m^{3}}
  = 1.89\times10^{-13}~\mathrm{kg}, \\[4pt]
  C_{\mathrm{Si}} &= \rho_{\mathrm{Si}} c_{p,\mathrm{Si}} V_{\mathrm{Si}}
  = 1.631\times10^{6}~\mathrm{J\,m^{-3}\,K^{-1}} \times 8.096\times10^{-17}~\mathrm{m^{3}}
  = 1.32\times10^{-10}~\mathrm{J\,K^{-1}}.
\end{align}

\paragraph{Pyrolytic-carbon pillars (per pillar).}
\begin{align}
  V_{\text{pillar}}
  &= \pi(1.5\times10^{-6}~\mathrm{m})^2 (3\times10^{-6}~\mathrm{m})
   = 2.12\times10^{-17}~\mathrm{m^{3}}, \\[4pt]
  m_{\text{pillar}} &= \rho_{\mathrm{pc}} V_{\text{pillar}}
  = 1900~\mathrm{kg\,m^{-3}} \times 2.12\times10^{-17}~\mathrm{m^{3}}
  = 4.03\times10^{-14}~\mathrm{kg}, \\[4pt]
  C_{\text{pillar}} &= \rho_{\mathrm{pc}} c_{p,\mathrm{pc}} V_{\text{pillar}}
  = 1.349\times10^{6}~\mathrm{J\,m^{-3}\,K^{-1}} \times 2.12\times10^{-17}~\mathrm{m^{3}}
  = 2.86\times10^{-11}~\mathrm{J\,K^{-1}}.
\end{align}

\paragraph{Total for six pillars.}
\begin{equation}
  C_{\text{pillars,tot}} = N\,C_{\text{pillar}} = 6 \times 2.86\times10^{-11}~\mathrm{J\,K^{-1}}
  = 1.72\times10^{-10}~\mathrm{J\,K^{-1}}.
\end{equation}

\subsection{Total heat capacity and composition}
\begin{equation}
  C_{th} = C_{\mathrm{Si}} + C_{\text{pillars,tot}}
  = 1.32\times10^{-10}~\mathrm{J\,K^{-1}} + 1.72\times10^{-10}~\mathrm{J\,K^{-1}}
  = 3.04\times10^{-10}~\mathrm{J\,K^{-1}}.
\end{equation}

This yields a breakdown by component of $43.5\%$ for the silicon membrane and  $56.5\%$ for the pillars.

\subsection{Assumptions}
\label{subsec:heat capacity - assumptions}

\begin{itemize}
    \item Zirconia is omitted by assumption as (i) zirconia is largely confined to the anchors/tethers or otherwise strongly thermalized to the bath so that its temperature participation (average $\Delta T$ relative to the pixel) is small, and (ii) its suspended volume is negligible compared with the Si membrane and carbon pillars. In the single-mode RC picture, only the heat capacity co-localized with the pixel’s temperature rise contributes appreciably to $C_{th}$; mass tightly coupled to the bath contributes little to the measured time constant.
    \item Thin surface layers or residues (e.g. resist) are not included, since these are typically orders of magnitude smaller in volume than the Si slab and pillars.
    \item Temperature-independent $c_p$ over the operating range. The calculation uses room-temperature $c_p$ values. For the small-signal temperature perturbations of the bolometer, the variation of $c_p$ is a higher-order correction and does not affect the leading-order estimate of $C_{th}$.

\end{itemize}

\subsection{Thermal time constant prediction}
\label{subsec:heat capacity - time constant pred}

Using the experimentally determined $R_{th} \approx 1.08\times10^{5}~\mathrm{K\,W^{-1}}$
and the calculated $C_{th} \approx 3.04\times10^{-10}~\mathrm{J\,K^{-1}}$,
we predict
\[
    \tau_{th}^{pred} \approx R_{th} C_{th} \approx 33~\mu\mathrm{s}.
\]
The close agreement between the measured ($\tau_{th}^{meas} \approx 27~\mu\mathrm{s}$) and estimated values confirms the internal consistency of the thermal model and the extracted parameters. We attribute the remaining discrepancy to uncertainty in the geometric and material parameters used in the estimate.

\section{Noise-budget analysis}
\label{sec:noise}

\subsection{Sequential noise characterization}
\label{subsec:noise - charact}

\begin{figure*}[!htbp]  % You can use [t], [b], [h], or [!htbp] for placement
    \centering
    \includegraphics[width=0.8\linewidth]{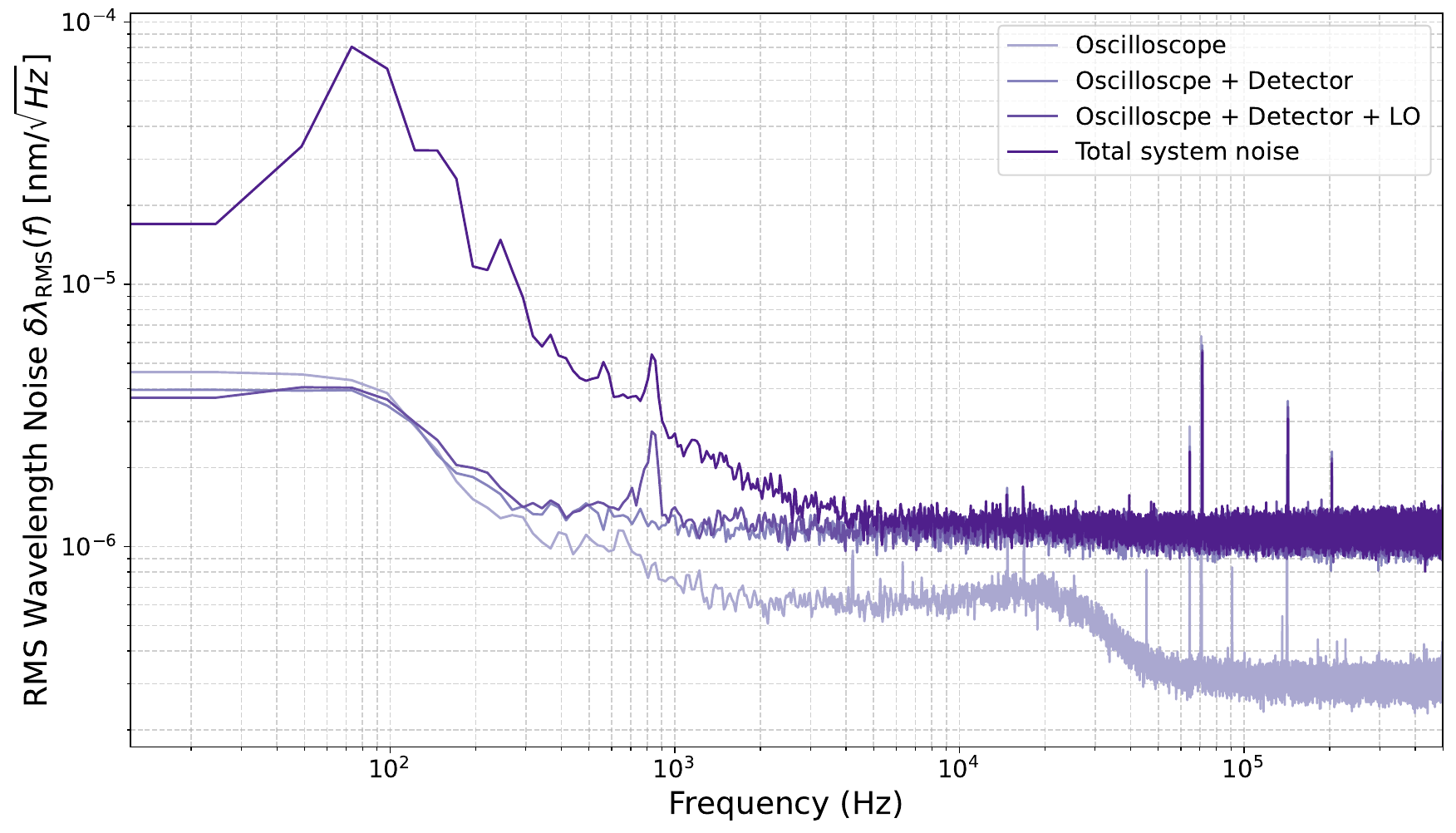} 
    \caption{\textbf{Sequential noise characterization of the optical readout.} RMS wavelength-noise amplitude spectral density $\delta\lambda_{\mathrm{RMS}}(f)$
    derived from continuous time-domain records using Welch’s method \cite{welch1967use} and the independently measured calibration slope $S$.
    Traces correspond to: (i) oscilloscope only (digitizer noise floor), (ii) oscilloscope~+~detector (electronics floor),
    (iii) oscilloscope~+~detector~+~LO with the signal blocked (adds LO shot and residual technical noise),
    and (iv) total system noise under full homodyne operation with the cavity aligned.
    At low frequencies, technical noise dominates before a white plateau is reached; the instrument-limited floor is
    $S_{\lambda} = 1.2~\mathrm{fm\,Hz^{-1/2}}$.}
    \label{fig:SI_noise_budget}
\end{figure*}

To quantify the different contributions to the readout noise, four spectra were recorded under identical conditions (Fig.~\ref{fig:SI_noise_budget}). The first trace, obtained with the oscilloscope input terminated, defines the digitizer noise floor. Adding the balanced detector in darkness raises this baseline to the electronics floor of the detector–TIA chain. Introducing the local oscillator (LO) while blocking the signal adds the LO shot noise and any residual technical noise that escapes common-mode rejection. The final trace, acquired with the cavity aligned and the lock set to the quadrature point used for responsivity calibration, represents the total system noise. At the moderate LO power used here, this spectrum coincides with the electronics-limited plateau, indicating that the current measurement is instrument-limited rather than shot-noise-limited. Operating at higher LO power would gradually lift the plateau toward the shot-noise regime, a condition intentionally avoided during calibration to prevent artefacts (see Sec.~\ref{subsec:perf scaling - rationale}).

\subsection{Voltage‑to‑wavelength calibration}
\label{subsec:noise - calibration}

The open‑loop sweep of the probe laser provides a direct voltage‑to‑wavelength calibration for the balanced‑homodyne readout. With the LO phase set to quadrature, the detector measures the signal‑field quadrature of the cavity reflection (Supplementary Fig.~\ref{fig:SI_setup}). For a single‑pole optical resonance the complex field response can be written
\[
\chi(\lambda)=\frac{1}{1+i\varepsilon},\qquad 
\varepsilon\equiv\frac{\lambda-\lambda_0}{\Gamma_\lambda},
\]
where $\lambda_0$ is the resonance wavelength and $\Gamma_\lambda$ is the half‑width at half‑maximum (HWHM) in wavelength units. The homodyne difference voltage then follows the familiar dispersive form
\[
v_h(\lambda)=v_\mathrm{off}-V_0\,\frac{\varepsilon}{1+\varepsilon^2},
\]
with $V_0$ an overall gain constant and $v_\mathrm{off}$ a slowly varying background. Defining the normalized Lorentzian intensity \(L(\lambda)=1/(1+\varepsilon^2)\), one finds
\[
\frac{dL}{d\lambda}=-\frac{2}{\Gamma_\lambda}\,\frac{\varepsilon}{(1+\varepsilon^2)^2}
\quad\Rightarrow\quad
v_h(\lambda)
=\frac{\Gamma_\lambda V_0}{2}\,(1+\varepsilon^2)\,\frac{dL}{d\lambda}+v_\mathrm{off}.
\]
Hence, in the vicinity of the lock point \(\lambda_{\mathrm{op}}\simeq\lambda_0\) where \(|\varepsilon|\ll1\), the measured lineshape is proportional to the \emph{derivative of a Lorentzian}. The local transduction slope used for calibration is
\[
S\;\equiv\;\left.\frac{dv}{d\lambda}\right|_{\lambda=\lambda_{\mathrm{op}}}
=\left.\frac{dv_h}{d\lambda}\right|_{\lambda_0}
=-\,\frac{V_0}{\Gamma_\lambda},
\qquad\text{so that}\qquad |S|=\frac{V_0}{\Gamma_\lambda}.
\]
This model is what we fit to the sweep in Fig.~3a (main text) to extract \(S\) at the point of maximum slope.\\

A long time-domain voltage record $v(t)$ is converted to the one-sided voltage power spectral density (PSD) $S_{v}(f)$ using Welch’s method \cite{welch1967use}. 
Using the independently measured slope at the quadrature lock point, we obtain the calibrated wavelength-noise PSD
\begin{equation}
    S_{\lambda}(f) = \frac{S_{v}(f)}{S^{2}}, 
    \qquad 
    \sqrt{S_{\lambda}(f)} = \frac{\sqrt{S_{v}(f)}}{|S|}.
    \tag{S4.1}
\end{equation}
The spectrum shows low-frequency technical noise before reaching a white plateau; its level defines the electronics-limited floor reported in the main text,
\[
    \sqrt{S_{\lambda}} = 1.2~\mathrm{fm\,Hz^{-1/2}}.
\]

\subsection{Thermo-refractive noise (TRN) limit}
\label{subsec:noise - TRN limit}

For a single thermal mode characterized by $R_{th}$ and $C_{th}$ (Sec.~\ref{sec:heat capacity}), the fluctuation--dissipation theorem \cite{callen_irreversibility_1951} yields the one-sided temperature-noise PSD
\begin{equation}
    S_{T}(f) = \frac{4 k_{B} T_{0}^{2} R_{th}}{1 + (2\pi f \tau_{th})^{2}}, 
    \qquad 
    \tau_{th} = R_{th} C_{th}.
\end{equation}

Temperature fluctuations shift the cavity resonance wavelength by
\begin{equation}
    \delta\lambda = \Lambda_{T} \, \delta T, 
    \qquad 
    \Lambda_{T} \equiv \lambda_{0}\!\left(\alpha + \frac{1}{n}\frac{dn}{dT}\right),
\end{equation}
where $\alpha$ is the linear thermal expansion coefficient and $dn/dT$ is silicon's thermo-optic coefficient (values in Supplementary Table~\ref{tab:parameters}).  
The TRN-limited wavelength-noise PSD \cite{levin_fluctuationdissipation_2008, PhysRevX.10.041046} is therefore
\begin{equation}
    S_{\lambda,\mathrm{TRN}}(f) = \Lambda_{T}^{2} S_{T}(f) 
    = \Lambda_{T}^{2} \frac{4 k_{B} T_{0}^{2} R_{th}}{1 + (2\pi f \tau_{th})^{2}}.
\end{equation}

In the low-frequency limit this approaches a white floor
\begin{equation}\label{eq:TRN_floor}
    \sqrt{S_{\lambda,\mathrm{TRN}}(0)} 
    = \Lambda_{T} \sqrt{4 k_{B} T_{0}^{2} R_{th}},
\end{equation}
set by the thermodynamics of the pixel.  
Evaluating Eq.~\ref{eq:TRN_floor} with the device parameters (Supplementary Table~\ref{tab:parameters} and Sec.~\ref{sec:heat capacity}) gives the TRN floor quoted in the main text,  
\[
    \sqrt{S_{\lambda,\mathrm{TRN}}} \approx 0.064~\mathrm{fm\,Hz^{-1/2}}.
\]
This is well below the measured electronics-limited plateau, confirming that the present readout masks the resonator’s fundamental limit. Further improvements in device engineering could reduce this gap, additional discussion is provided in Section~\ref{sec:perf scaling}.

\section{Calibration robustness and performance scaling}
\label{sec:perf scaling}

\subsection{Rationale and conventions}
\label{subsec:perf scaling - rationale}

In the main text we calibrated and characterized the homodyne readout at moderate local-oscillator (LO) power to maintain a clean derivative-of-Lorentzian calibration curve; 
at higher LO powers, weak back-reflections made the lineshape determination less reliable. 
Here we analyze how improvements to the photonic-crystal cavity. Specifically, higher optical quality factor $Q$ and higher collection (zero-order) efficiency $\eta$, 
the fraction of input power coherently collected into the detected mode affect 
(i) the transduction gain $S$, 
(ii) the input-referred readout noise (electronics- and shot-noise floors), 
and (iii) the feasibility of operating at the TRN limit. 

We adopt the single-pole cavity model measured in reflection at the quadrature lock point, 
use one-sided PSDs, and report ASDs in $\mathrm{fm}/\sqrt{\mathrm{Hz}}$. 
Throughout, $P_\mathrm{in}$ denotes the optical power delivered to the device, 
and $P_\mathrm{sig} = \eta P_\mathrm{in}$ the collected signal power.

\subsection{Homodyne transduction: slope vs. device parameters}
\label{subsec:perf scaling - device param}

Near the operating point the balanced-homodyne output is linear in wavelength,
\begin{equation}
v \approx v_0 + S\,\delta\lambda,
\qquad\text{(slope $S$ in V/nm).}
\end{equation}

For a single-port cavity probed in reflection and read out at the quadrature point, the difference current is
\begin{equation}
i_h \approx 2\,\mathcal{R}\,\sqrt{P_{\mathrm{LO}} P_{\mathrm{sig}}}\,\sin\varphi,
\end{equation}
where $\mathcal{R}$ is the photodiode responsivity (A/W), 
$P_{\mathrm{LO}}$ is the LO power at the detector, 
$P_{\mathrm{sig}} = \eta P_{\mathrm{in}}$ is the collected signal power, 
and $\varphi$ is the signal-field phase relative to the LO at the detection port.

Linearizing at quadrature ($\sin\varphi \simeq \varphi$) and converting current to voltage with the transimpedance $R_{\mathrm{TIA}}$ gives the calibration slope
\begin{equation}
S \;=\; \frac{dv}{d\lambda}
= 2\,\mathcal{R}\,R_{\mathrm{TIA}}\,\sqrt{P_{\mathrm{LO}}\eta P_{\mathrm{in}}}
\left|\frac{d\varphi}{d\lambda}\right|.
\label{eq:S_supp}
\end{equation}

For a single-pole resonance the phase derivative at the optimum lock point is proportional to $Q/\lambda_0$, 
up to an order-unity lineshape factor $\zeta$ (accounting for exact coupling and lock point):
\begin{equation}
\left|\frac{d\varphi}{d\lambda}\right| = \zeta\,\frac{2Q}{\lambda_0},
\qquad 0 < \zeta \le 1.
\label{eq:phase_slope_supp}
\end{equation}

Collecting constants into $\kappa_0 \equiv 4\zeta\,\mathcal{R}R_{\mathrm{TIA}}$, we obtain the following scaling
\begin{equation}
S = \kappa_0\,\frac{Q}{\lambda_0}\,
\sqrt{\eta P_{\mathrm{in}} P_{\mathrm{LO}}}
\quad\Longrightarrow\quad
S \propto Q\,\eta^{1/2}
\quad\text{(for fixed $P_\mathrm{in}$, $P_\mathrm{LO}$, electronics).}
\label{eq:S_scaling_supp}
\end{equation}

This shows that a higher-$Q$ cavity produces a steeper phase response per nanometer, and better collection increases the interferometric signal: both raise the transduction gain $S$.

\subsection{Input-referred readout noise - closed forms}
\label{subsec:perf scaling - readout noise}

Noise is reported in wavelength units by referring the measured voltage PSD $S_v$ through the local slope (see Sec.~\ref{subsec:noise - calibration}):
\begin{equation}
S_\lambda \;=\; \frac{S_v}{S^2},
\qquad\text{hence}\qquad
\sqrt{S_\lambda} \;=\; \frac{\sqrt{S_v}}{|S|}.
\label{eq:refer_rule}
\end{equation}

\textit{Electronics-limited floor.}
Let $\sqrt{S_{v,\mathrm{elec}}}$ be the detector/TIA voltage ASD (intrinsic to the electronics). 
Referring to wavelength,
\begin{equation}
\sqrt{S_{\lambda,\mathrm{elec}}}
= \frac{\sqrt{S_{v,\mathrm{elec}}}}{S}
= \frac{\lambda_0}{Q\,\sqrt{\eta P_\mathrm{in}}}\;
\underbrace{\frac{\sqrt{S_{v,\mathrm{elec}}}}{\kappa_0\,\sqrt{P_\mathrm{LO}}}}_{\displaystyle C_\mathrm{elec}}
\label{eq:elec_floor_s53}
\end{equation}
So increasing $Q$ and $\eta$ lowers the equivalent input electronics floor via a larger slope $S$.\\

\textit{Shot-noise-limited floor.}
The ouput shot noise is 
$\sqrt{S_{v,\mathrm{shot}}}=R_\mathrm{TIA}\sqrt{2q\,\mathcal R\,P_\mathrm{LO}}$.
Referring to wavelength and substituting $S$ cancels the LO-dependence:
\begin{equation}
\sqrt{S_{\lambda,\mathrm{shot}}}
= \frac{R_\mathrm{TIA}\sqrt{2q\,\mathcal R\,P_\mathrm{LO}}}{S}
= \frac{\lambda_0}{Q\,\sqrt{\eta P_\mathrm{in}}}\;
\underbrace{\frac{\sqrt{2q/\mathcal R}}{2\zeta}}_{\displaystyle C_\mathrm{shot}}.
\label{eq:shot_floor_s53}
\end{equation}
Once shot-noise limited, only $Q$ and $\eta$ (and $P_\mathrm{in}$) lower the input-referred shot floor.\\

\textit{Total readout floor.}
It is convenient to collect contributions as
\begin{equation}
\sqrt{S_{\lambda,\mathrm{readout}}}
= \frac{\lambda_0}{Q\,\sqrt{\eta P_\mathrm{in}}}\;
\Big(C_\mathrm{elec}^2 + C_\mathrm{shot}^2 + C_\mathrm{tech}^2\Big)^{1/2}
\label{eq:readout_master_s53}
\end{equation}
where $C_\mathrm{tech}$ accounts for any residual LO technical noise that is not perfectly common-mode rejected.
Equation \eqref{eq:readout_master_s53} explicitly identifies the governing design parameters. An increase in either the cavity quality factor $Q$ or the collection efficiency $\eta$ reduces the equivalent input noise in proportion to $1/(Q \sqrt{\eta})$, thereby improving the overall readout sensitivity.

\subsection{Condition for operating at the TRN limit}
\label{subsec:perf scaling - condition for TRN limit}

The TRN floor is determined by the device’s thermodynamics and thermo-optic coupling (Sec.~\ref{subsec:noise - TRN limit}) and does not depend on $Q$ or $\eta$. 
Operating at the TRN limit requires
\begin{equation}
\sqrt{S_{\lambda,\mathrm{readout}}} \;\le\; \sqrt{S_{\lambda,\mathrm{TRN}}}.
\label{eq:TRN_inequality}
\end{equation}

Substituting Eq.~\eqref{eq:readout_master_s53} into \eqref{eq:TRN_inequality} yields the following sufficiency condition:
\begin{equation}
Q\,\sqrt{\eta}
\;\ge\;
\frac{\lambda_0}{\sqrt{P_\mathrm{in}}}\,
\frac{\big(C_\mathrm{elec}^{2}+C_\mathrm{shot}^{2}+C_\mathrm{tech}^{2}\big)^{1/2}}
{\sqrt{S_{\lambda,\mathrm{TRN}}}}.
\label{eq:TRN_condition_formal}
\end{equation}

With the present device, the measured white plateau is 
$1.2~\mathrm{fm}/\sqrt{\mathrm{Hz}}$, 
whereas the calculated TRN limit is 
$0.064~\mathrm{fm}/\sqrt{\mathrm{Hz}}$,
corresponding to a gap of approximately $19\times$ in amplitude.
Because $\sqrt{S_{\lambda,\mathrm{readout}}} \propto 1/(Q\sqrt{\eta})$, 
any combination of design improvements that increases $Q\sqrt{\eta}$ by roughly a factor of~19 
(or equivalently reduces $C_\mathrm{elec}$ by the same factor) 
is sufficient to reach the TRN floor. 
Beyond this point, further increases in $Q$ or $\eta$ do not lower the measured noise, 
and the spectrum saturates at the TRN level unless the material or thermal parameters of the device are altered \cite{PhysRevX.10.041046}, showing one more interesting avenue to explore.\\

Moreover, since we saw that $S \propto Q \eta^{1/2} \sqrt{P_{LO}}$, improving these device metrics also enables to reach the same calibration precision as the one used in the main text with much lower LO power. This would directly erase the back-reflection sensitivity that motivated moderate-LO operation.

%%% GLOBAL TABLE %%%

\begin{table}[h!]
\centering
\caption{Nomenclature and parameter summary}
\label{tab:parameters}
\begin{tabular}{llcl}
\hline
\textbf{Symbol} & \textbf{Quantity} & \textbf{Value} & \textbf{Reference / Source} \\
\hline
$T_0$ & Ambient temperature & $295~\mathrm{K}$ & Experimental condition \\
$\rho_{\mathrm{Si}}$ & Silicon mass density & $2330~\mathrm{kg\,m^{-3}}$ & \cite{arblaster_selected_2018} \\
$c_{p,\mathrm{Si}}$ & Silicon specific heat capacity & $700~\mathrm{J\,kg^{-1}\,K^{-1}}$ & \cite{chase_nist-janaf_1998} \\
$\alpha_{\mathrm{Si}}$ & Silicon linear thermal expansion coeff. & $2.6\times10^{-6}~\mathrm{K^{-1}}$ & \cite{swenson_recommended_1983} \\
$n_{\mathrm{Si}}$ & Silicon refractive index & $3.48$ & \cite{komma_thermo-optic_2012} \\
$\left.\dfrac{dn}{dT}\right|_{\mathrm{Si},\,1550\,\mathrm{nm}}$ & Silicon thermo-optic coefficient & $1.86\times10^{-4}~\mathrm{K^{-1}}$ & \cite{komma_thermo-optic_2012} \\
$\rho_{\mathrm{pc}}$ & Pyrolytic-carbon mass density & $1.90\times10^{3}~\mathrm{kg\,m^{-3}}$ & \cite{pierson_handbook_1993} \\
$c_{p,\mathrm{pc}}$ & Pyrolytic-carbon specific heat capacity & $710~\mathrm{J\,kg^{-1}\,K^{-1}}$ & \cite{chase_nist-janaf_1998} \\
$A_{\mathrm{Si}}$ & Silicon membrane area & $23~\mu\mathrm{m} \times 16~\mu\mathrm{m}$ & Device design \\
$t_{\mathrm{Si}}$ & Silicon membrane thickness & $220~\mathrm{nm}$ & Process specification \\
$r_{\mathrm{pillar}}$ & Pillar radius & $1.5~\mu\mathrm{m}$ & SEM estimation \\
$h_{\mathrm{pillar}}$ & Pillar height & $3~\mu\mathrm{m}$ & SEM estimation \\
$N_{\mathrm{pillar}}$ & Number of pillars & $6$ & Device layout \\
$\lambda_0$ & Cavity resonance wavelength & $1568.83~\mathrm{nm}$ & Measured (this work) \\
$Q$ & Cavity quality factor & $3.9 \times 10^4$ & Measured (this work) \\
$R_{\mathrm{th}}$ & Effective thermal resistance (device) & $1.08\times10^{5}~\mathrm{K\,W^{-1}}$ & Measured (this work) \\
\hline
\end{tabular}
\end{table}

\end{document}